\pdfoutput=1
\documentclass[twocolumn]{aastex701}
\newcommand{\dd}{\mathrm{d}}
\newcommand{\pneo}{{\tt pneo}\xspace} 

\usepackage{amsmath}
\graphicspath{{Figures/}{./}{../}}
\usepackage{xcolor}
\usepackage{xspace}

\begin{document}

\title{A Bayesian Approach to Prioritizing the Follow-Up of Near-Earth Objects}

\correspondingauthor{Devanshi Singh}
\email{ds2004@uw.edu}

\author[orcid=0009-0004-0059-3229]{Devanshi Singh}
\affiliation{DiRAC Institute and the Department of Astronomy, University of Washington, 3910 15th Ave NE, Seattle, WA, 98195, U.S.A.}
\email{ds2004@uw.edu}

\author[orcid=0000-0003-1996-9252]{Mario Juri\'{c}}
\affiliation{DiRAC Institute and the Department of Astronomy, University of Washington, 3910 15th Ave NE, Seattle, WA, 98195, U.S.A.}
\email{mjuric@astro.washington.edu}

\author[orcid=0000-0001-5250-2633]{\v{Z}eljko Ivezi\'{c}}
\affiliation{DiRAC Institute and the Department of Astronomy, University of Washington, 3910 15th Ave NE, Seattle, WA, 98195, U.S.A.}
\email{ivezic@astro.washington.edu}

\begin{abstract}

We present \pneo, a Bayesian classifier that computes the probability that a single-night tracklet belongs to a near-Earth object (NEO), using only its sky position, angular motion, and apparent magnitude. A tracklet does not determine the object's distance or radial velocity, so \pneo integrates a model of the Solar System's small-body populations over all bound orbits consistent with it. The result is the exact posterior probability under that model, with no tuned scores or separately chosen priors: its classifications are optimal for the adopted model, and its probabilities are calibrated.

On real discovery tracklets from the NEO Confirmation Page, the classifier is well calibrated and its selection is $1.14\times$ purer than that of digest2, the classifier now in use, at the same completeness. The advantage grows at Rubin's depth, where the faint background of main-belt and more distant objects is still largely undiscovered. On simulated tracklets to that depth, its selection is $2\times$ purer than that of digest2 (96\% against 48\%). On the Rubin alert stream and Rubin Data Preview~2 its selection is 1.4--$1.7\times$ purer than that of digest2, and its predicted NEO counts are close to those observed. By raising the fraction of follow-up targets that are NEOs, \pneo could potentially double the effective capacity of NEO follow-up in the Rubin era.

\end{abstract}


\section{Introduction}
\label{sec:intro}

Near-Earth Objects (NEOs)---asteroids and comets with orbits that bring them
within 1.3~AU of the Sun---are typically discovered by wide-field sky surveys
that detect new moving objects by taking a sequence of two to four images
within a single night. Detections that cannot be matched to any known body are
reported to the Minor Planet Center (MPC) as a \emph{tracklet}. Candidates
consistent with an NEO orbit are identified and posted to the MPC's NEO
Confirmation Page \citep[NEOCP;][]{marsden1998,spoto2021}. A candidate remains
there until follow-up astrometry either secures its orbit---earning it a
designation---or identifies it as a main-belt asteroid or other non-NEO. In
this way, the NEOCP serves as the central community clearinghouse for
collecting NEO candidates and prioritizing their follow-up.

The Vera C. Rubin Observatory's Legacy Survey of Space and Time
\citep[LSST;][]{ivezic2019} will make this prioritization much harder. Today,
roughly 5--27 candidates are added to the NEOCP each night, depending on lunar
phase \citep{wagg2025}. But simulating the first year of LSST under the current
posting criterion, a digest2 score of at least 65, \citet{wagg2025} estimated
that Rubin alone could submit about 129 candidates per night, around eight
times the present rate, with only 8.3\% of the candidates being NEOs. Most of
the remainder are faint, previously undiscovered main-belt asteroids. Even
removing candidates that Rubin is likely to recover on its own reduces the
list only to about 64 per night, with the NEO fraction still below 10\%.

Such an increase presents an immediate challenge to NEO follow-up efforts'
limited resources. To mitigate it, \citet{wagg2025} propose several options,
each designed to make the LSST-era follow-up highly selective (including
additional cuts on ecliptic longitude/latitude, and delaying rapid Rubin
candidate follow-up until a few years into the survey, once the background of
unseen $V<24$~mag asteroids is reduced by discovery). They do not, however,
examine the efficiency of, or possible improvements to, the criterion for
NEOCP placement itself: the digest2 score.

The tool currently used to prioritize tracklets for the NEOCP is digest2
\citep{keys2019}, recently extended to read ADES astrometry
\citep{veres2023}. For each tracklet, digest2 samples possible topocentric
distances, constructs trial orbits, and compares them with a population model
divided into orbital classes. It returns a score from 0 to 100; candidates are
normally posted when the NEO score, computed using the undiscovered
(``NoID'') population, reaches 65, subject to MPC review
\citep{veres2025ml}.

digest2, while undoubtedly successful and widely used over the past decade, is
not optimal. First, it is not a Bayesian classifier and does not produce
calibrated posterior probabilities; because Bayesian classifiers achieve the
minimum attainable misclassification rate for any class-conditional model and
prior \citep[e.g.,][]{duda2001}, rebuilding a classifier on a Bayesian
foundation represents an opportunity for a more principled approach, one that
could lead to better classification performance.

Second, digest2 is not calibrated: the score it produces does not correspond
to the probability that an object is an NEO; \citet{keys2019} describe it as a
``pseudo-probability.'' A classifier producing a calibrated $P(\mathrm{NEO})$
would have a direct interpretation: among tracklets assigned probability $p$,
a fraction $p$ should be NEOs. Follow-up programs could then choose thresholds
based on the expected NEO yield and contamination, rather than calibrating a
ranking score empirically.

This paper aims to address both of these issues by introducing \pneo, a
Bayesian classifier that estimates the posterior probability of NEO
membership, $P(\mathrm{NEO})$, directly from a discovery tracklet and its
apparent brightness, using an absolute model of the orbital and size
distributions of the NEO and background populations. We show that \pneo is
well calibrated: on 800\,000 of our simulated tracklets it predicts 10\,543
NEOs, against 10\,550 observed. Secondly, it is efficient: at the completeness
of the current NEOCP posting threshold, its selection is 96\% pure, compared
with 48\% for digest2, mainly because it rejects the distant objects that
digest2 mistakes for NEOs. On real discovery tracklets from the NEOCP it
selects about 40\% fewer non-NEOs than digest2 at comparable completeness, and
on the Rubin alert stream 50--75\% fewer. Using this classifier as a tool to
prioritize NEO follow-up could significantly increase the efficiency of NEO
follow-up in the Rubin era.

This paper is organized as follows. In Section~\ref{sec:methods} we derive
$P(\mathrm{NEO})$ for a single-night tracklet and its brightness, including
the change of measure from the ranging coordinates to orbital elements.
Section~\ref{sec:population} describes the population model (deliberately kept
the same as the digest2 model), and Section~\ref{sec:implementation} the
implementation. Section~\ref{sec:validation} validates \pneo on simulated
tracklets, testing its calibration, comparing it with digest2, and testing
its robustness to astrometric noise. Section~\ref{sec:realdata} applies the
classifier to real data from the NEOCP, the Rubin alert stream, and Rubin Data
Preview~2. We
discuss the implications for LSST-era follow-up, and the limitations of the
method, in Section~\ref{sec:discussion}, and summarize in
Section~\ref{sec:summary}.

\section{Estimating p(NEO): A Bayesian tracklet classifier}
\label{sec:methods}

A single-night tracklet carries information about the object's on-sky position and motion, but not the topocentric distance (range) and range rate. It is range the separates a
small, nearby NEO from a larger, distant asteroid with the same position,
motion and apparent brightness, and the tracklet does not measure it. Similarly,
its magnitude is apparent, and does not carry the information about the size.

Absent that information, but knowing the dynamical properties of 
the overall population of small bodies -- size and dynamical distribution -- the next best thing one can do is to consider all allowable distance the object could have, and weigh
how plausible each one is. This statistical ranging \citep[][e.g.]{virtanen2001,muinonen2006,chesley2005,farnocchia2015,spoto2018} gives
us a way to compute the posterior probability for each tracklet to be
at a certain range and range rate.

These conditional probabilities are computed for all combinations of (range, range rate, and absolute magnitude), and summed (marginalized) over those three unknowns: once conditioned on $q < 1.3$ (for NEOs), and once for all objects. The probability of a tracklet belonging to an NEO, $P(NEO)$ is then simply the ratio of the two. This is \pneo in a nutshell.
\medskip

While conceptually simple, each of these steps carries important implementation details. We go through them in the subsections to follow.

\subsection{Observables and the two unknowns}
\label{sec:observables}

For each candidate, a survey reports a tracklet consisting of two or more
detections of the same moving source within one night. Each detection provides
a right ascension $\alpha$, declination $\delta$, observation time $t$, and
an apparent magnitude in the survey's photometric band (for LSST, one of
$u,g,r,i,z,y$). We assume linear motion over the tracklet and fit
$\alpha(t)$ and $\delta(t)$ to estimate the position and angular rates
$\dot\alpha$ and $\dot\delta$ at the epoch of the first detection. We define the attributable
$A=(\alpha,\delta,\dot\alpha,\dot\delta)$ \citep{milani2004}, where
$\dot\alpha=d\alpha/dt$ is the rate of change of the right-ascension
coordinate itself rather than the on-sky rate $\dot\alpha\cos\delta$.
The photometry is converted to a single Johnson $V$ magnitude
(Section~\ref{sec:photometry}), while the
observer's heliocentric position $\mathbf{R}$ and velocity
$\dot{\mathbf{R}}$ are computed from the JPL DE440s planetary ephemeris
\citep{park2021} and the observatory's location, taken from the MPC list of
observatory codes.

\subsection{From a hypothesis to an orbit}
\label{sec:geometry}

A short tracklet does not determine the object's topocentric distance
$\rho$ or range-rate $\dot\rho$. Different choices of
$(\rho,\dot\rho)$ can therefore reproduce the same attributable while
corresponding to different heliocentric orbits. We call each particular
choice of $(\rho,\dot\rho)$ a \emph{range hypothesis}, or simply a
hypothesis.

Let
$\hat{\mathbf{u}}=(\cos\delta\cos\alpha,\cos\delta\sin\alpha,\sin\delta)$
be the line-of-sight unit vector, with
$\hat{\mathbf{u}}_\alpha=\partial\hat{\mathbf{u}}/\partial\alpha$ and
$\hat{\mathbf{u}}_\delta=\partial\hat{\mathbf{u}}/\partial\delta$. The
heliocentric position and velocity implied by a hypothesis are
\begin{align}
\mathbf{r} &= \mathbf{R} + \rho\,\hat{\mathbf{u}}, \label{eq:r}\\
\mathbf{v} &= \dot{\mathbf{R}} + \dot{\rho}\,\hat{\mathbf{u}}
+ \rho\left(\dot{\alpha}\,\hat{\mathbf{u}}_\alpha +
\dot{\delta}\,\hat{\mathbf{u}}_\delta\right). \label{eq:v}
\end{align}

Standard two-body relations then give the osculating heliocentric ecliptic elements at the tracklet epoch: semimajor axis $a$, eccentricity $e$,
inclination $i$, longitude of the ascending node $\Omega$, argument of
perihelion $\omega$, and mean anomaly $M$. We denote these collectively by
$E=(a,e,i,\Omega,\omega,M)$. Together, Equations~(\ref{eq:r}) and
(\ref{eq:v}) define the map
$\Phi:(A,\rho,\dot\rho)\mapsto E$, so each hypothesis corresponds to one possible, bound
trial orbit consistent with the tracklet. Hypotheses that imply unbound heliocentric orbits are excluded
(Section~\ref{sec:ar}). Both population components contain only bound
orbits, so the prior density is zero for these hypotheses and excluding them does
not change $P(\mathrm{NEO})$; it also keeps $\Phi$ defined, since $a$ and
$M$ are not defined for $e\geq1$.

\subsection{Photometry}
\label{sec:photometry}

Each hypothesis specifies a trial orbit, but evaluating it
against the population model also requires its absolute magnitude $H$. The
observed apparent magnitude provides this additional constraint because the
implied $H$ depends on the distance and viewing geometry of the hypothesis.

We convert each detection magnitude to Johnson $V$ using the fixed band
offsets adopted by digest2 \citep{keys2019}, and use the median converted
magnitude for the tracklet. For each hypothesis, the heliocentric distance
$r=|\mathbf{r}|$ and phase angle $\alpha_{\rm ph}$, defined by the
Sun--object--observer geometry, follow from Equation~(\ref{eq:r}). Using the
$H,G$ phase function \citep{1989aste.conf..524B}, the implied absolute magnitude is
\begin{equation}
H(V,\rho) = V - 5\log_{10}(\rho r) + 2.5\log_{10}\Phi_{HG}(\alpha_{\rm ph};G),
\label{eq:H}
\end{equation}
with $G=0.15$ for all objects. Thus the same apparent magnitude corresponds
to a larger $H$ for a nearby hypothesis and a smaller $H$ for a distant one; for a
fixed albedo, these correspond to smaller and larger objects, respectively.

\subsection{Population densities and the marginal likelihood}
\label{sec:likelihood}

We divide the population into two classes,
$C\in\{\mathrm{NEO},\mathrm{other}\}$, with NEOs defined by
$q=a(1-e)<1.3$~AU. Each class is represented by an intrinsic number density
$n_C(E,H)$. By intrinsic, we mean the underlying Solar System population
rather than the subset of objects already discovered. The densities retain
their absolute normalization, so integrating $n_C$ over the full model
returns the total number of objects in that class. The relative abundance of
NEOs and non-NEOs is therefore already contained in the population model
(Section~\ref{sec:posterior}).

Here, both densities are derived from the S3M-based model used by digest2
(Section~\ref{sec:population}). The model is tabulated in $(q,e,i,H)$ and
assumes uniform $\Omega$, $\omega$, and $M$. This is an approximation:
NEOs show modest departures from uniformity in $\Omega$ and $\omega$
\citep{jeongahn2014} but we keep it at this stage to make comparisons to digest2 more straightforward; improvements to the population model are left
for future work. Outside the adopted model support, $n_C=0$ (Section~\ref{sec:truncation}).

To count the class-$C$ objects consistent with a tracklet, we evaluate
$n_C$ for each range hypothesis (Section~\ref{sec:geometry}) and sum over
them. The hypotheses lie on a grid in the observed coordinates
$(A,\rho,\dot\rho,V)$ (in practice, a grid in $(\rho,\dot\rho)$ at the
measured $A$ and $V$) rather than in orbital elements. Equal-sized cells in
this grid generally map to different volumes of element space, so the
population density must be multiplied by the appropriate change-of-variables
factor. For a small cell around one grid point, the expected number of
class-$C$ objects is

\begin{equation}
 \dd N_C = n_C(E,H)\,
\left|\det\frac{\partial E}{\partial(A,\rho,\dot\rho)}\right|\,
 \dd A\,\dd\rho\,\dd\dot\rho\,\dd V .
\label{eq:dN}
\end{equation}

The determinant (Section~\ref{sec:jacobian}) converts the grid volume into orbital-element
volume. The transformation from $V$ to $H$ contributes no additional factor
because $\partial H/\partial V=1$.

Summing over the admissible region $\mathcal{AR}$ of bound orbits
(Section~\ref{sec:ar}) gives the expected number of objects in a class:
\begin{multline}
\Lambda_C(A,V) = \iint_{\mathcal{AR}}
n_C\bigl(\Phi(A,\rho,\dot\rho),H(V,\rho)\bigr)\\
\times\left|\det\frac{\partial E}{\partial(A,\rho,\dot\rho)}\right|
\dd\rho\,\dd\dot\rho ,
\label{eq:lambda}
\end{multline}
which we evaluate numerically as
$\Lambda_C\approx\sum_k n_C(E_k,H_k)\,|J_k|\,w_k$ over grid points $k$ with
cell areas $w_k$.

$\Lambda_C$ is the class marginal likelihood. It measures how
much of the class-$C$ population is consistent with the observed tracklet
after marginalizing over distance and range-rate. More precisely, it is a
differential number density in observable space: the expected number of
objects per unit $\dd A\,\dd V$ around the measured values.

\subsection{The change of measure}
\label{sec:jacobian}

The determinant in Equation~(\ref{eq:lambda}) is simply the Jacobian of the change
of variables from the observed coordinates $(A,\rho,\dot\rho)$ to the
orbital elements $E$. In our case, it can be factored through the
heliocentric state vector,
$\partial E/\partial(A,\rho,\dot\rho)=
[\partial E/\partial(\mathbf{r},\mathbf{v})]\,
[\partial(\mathbf{r},\mathbf{v})/\partial(A,\rho,\dot\rho)]$,
so its determinant is the product of the determinants of the two factors.
Both have exact closed forms, derived in
Appendix~\ref{app:jacobian}. The transformation from observables to
Cartesian state contributes $\rho^4\cos^2\delta$, while the transformation
from Cartesian state to orbital elements contributes
$2/(\mu^{3/2}\sqrt{a}\,e\sin i)$, where $\mu$ is the Sun's gravitational
parameter. Because our population density is tabulated in $q$ rather than
$a$, we express it in those same coordinates, introducing
the additional factor $\partial q/\partial a=1-e$:
\begin{align}
&\left|\det\frac{\partial(q,e,i,\Omega,\omega,M)}
{\partial(A,\rho,\dot\rho)}\right|
\nonumber\\
&\qquad = \rho^4\cos^2\delta\,
\frac{2(1-e)}{\mu^{3/2}\sqrt{a}\,e\sin i}.
\label{eq:jacqei}
\end{align}

The $\rho^4$ factor can be understood as a phase-space volume effect. At
distance $\rho$, a fixed angular patch on the sky corresponds to a physical
area proportional to $\rho^2$. The same is true in velocity space: a fixed
range of angular rates corresponds to transverse velocities proportional to
$\rho$, giving another factor of $\rho^2$. Thus a small cell in the observed
position and motion represents a phase-space volume proportional to
$\rho^4$. Doubling the distance increases that volume by a factor of 16.
Without this weighting, nearby and distant hypotheses would be treated as
representing equal volumes, which gives too much relative weight to nearby,
often NEO-like solutions. digest2, which applies no change of measure, is
affected in exactly this way: the slow tracklets of distant objects admit
nearby, NEO-like solutions, and without the $\rho^4$ weighting these
dominate, producing most of digest2's false positives
(Sections~\ref{sec:premise} and~\ref{sec:disagreement}).

The remaining factors account for the orbital
coordinate system: $(1-e)$ follows from $q=a(1-e)$, while
$1/(e\sin i)$ is an integrable coordinate singularity rather than a physical
concentration of objects.

\subsection{The posterior}
\label{sec:posterior}

Since the two populations do not overlap, their marginal
likelihoods can be normalized to obtain the posterior probability of NEO
membership:
\begin{equation}
P(\mathrm{NEO}\mid A,V)
= \frac{\Lambda_{\rm NEO}}
{\Lambda_{\rm NEO}+\Lambda_{\rm other}}.
\label{eq:posterior}
\end{equation}

Given the population model, Equation~(\ref{eq:posterior}) is not an
approximation or a heuristic score but the {\em exact} posterior probability that follows
from Bayes' theorem: the population model supplies both the prior abundance
of each class and the likelihood of each orbital hypothesis, and the change
of measure (Section~\ref{sec:jacobian}) is required to express that
likelihood in the observed coordinates. And calibration is ensured by construction: if the population model
approximates the real sky well, the resulting probabilities will be calibrated: among a
large set of tracklets assigned $P(\mathrm{NEO})=0.8$, approximately 80\%
should be NEOs.

The remaining approximations lie in
the inputs (the population model and the treatment of the attributable as
exact) and in the numerical quadrature, which we discuss below and test in
Section~\ref{sec:validation}.

\subsection{Admissible region limits and quadrature}
\label{sec:ar}

The integral in Equation~(\ref{eq:lambda}) runs over the admissible region
$\mathcal{AR}$ \citep{milani2004}, the set of range hypotheses
$(\rho,\dot\rho)$ that produce a bound heliocentric orbit,
$\tfrac12|\mathbf{v}|^2-\mu/r<0$. We further restrict it to
$0.01\le\rho\le100$~AU.

The lower limit is roughly the radius of Earth's Hill sphere, inside which
Earth's gravity dominates and a heliocentric two-body orbit is not a
meaningful description; the upper limit is the outer perihelion edge of the
adopted population model, beyond which $n_C=0$ for nearly all hypotheses
(see below for the rare exceptions). Both limits
are configurable. Hyperbolic solutions are
excluded because interstellar objects are outside the scope of this work and
both population components contain only bound Solar System orbits. Our range
hypotheses thus play the role of the virtual asteroids used to sample this
region \citep{milani2004}.

Objects beyond the upper limit are not represented. In simulations
(Section~\ref{sec:gateresults}), distant objects with $q>30$~AU and
$H<4$ whose true distance exceeds 100~AU receive a spurious
$P(\mathrm{NEO})=1$, because their true solution lies outside the domain.
Such objects are rare and easy to recognize: at 100~AU, Earth's orbital
motion produces at most about $1.5''\,\mathrm{h}^{-1}$ of apparent motion,
so slow tracklets assigned $P(\mathrm{NEO})=1$ should be treated as outside
the model's domain.

\begin{figure*}
\centering
\includegraphics[width=\textwidth]{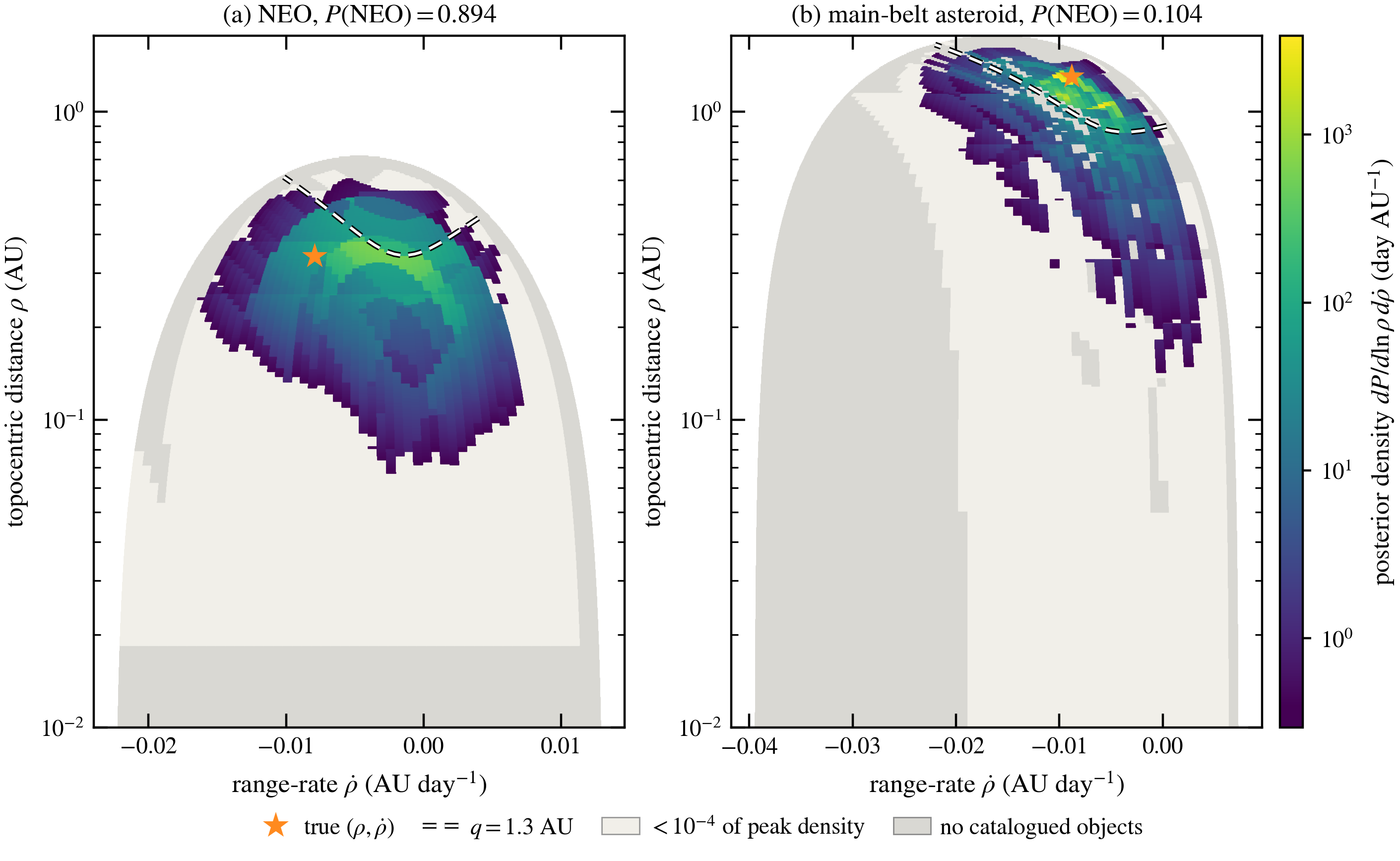}
\caption{Posterior over the hypothesis space for two tracklets from the self-consistency sample
(Section~\ref{sec:selfconsistency}), re-observed from the Rubin Observatory site (X05):
(a) an NEO ($a=2.35$~AU, $e=0.54$, $q=1.08$~AU, $H=23.3$) and
(b) a main-belt asteroid ($a=2.88$~AU, $e=0.35$, $i=37.8^\circ$, $q=1.87$~AU, $H=16.6$).
Each panel shows the admissible region, every topocentric distance $\rho$ and range-rate $\dot\rho$
consistent with the tracklet that gives a bound heliocentric orbit, drawn with one cell per range
hypothesis on the $1200\times50$ quadrature grid.
Color is the posterior density per unit $\ln\rho$ and $\dot\rho$, on one scale for both panels, so
color times displayed area is probability.
Pale cells hold less than $10^{-4}$ of the peak density; gray cells imply orbits or absolute
magnitudes outside the population model, where $n_C=0$ (in panel a this includes the band at
$\rho<0.018$~AU, where $H>30$).
The dashed curve is $q=1.3$~AU: cells below it imply NEO orbits.
The star marks the true $(\rho,\dot\rho)$.
The classifier gives $P(\mathrm{NEO})=0.894$ for the NEO and $0.104$ for the main-belt asteroid; in both, 97\%
of the posterior lies within a factor of two of the true distance.}
\label{fig:hypothesis}
\end{figure*}

Figure~\ref{fig:hypothesis} shows the resulting posterior over the
admissible region for one NEO and one main-belt tracklet. Three features
are worth noting. First, although each tracklet admits a wide range of
distances, nearly two decades in $\rho$, the posterior is concentrated in
a small part of the admissible region, close to the true distance and
range-rate: most hypotheses are implausible once the population model and
the change of measure are applied. Second, $P(\mathrm{NEO})$ is simply the
fraction of the posterior lying below the $q=1.3$~AU curve. Third, that
curve cuts through the posterior in both cases, so neither tracklet is
unambiguous. The classifier reflects this by assigning probabilities of
0.89 and 0.10, rather than 1 and 0, in proportion to the plausible
solutions on each side of the boundary.

\subsection{Extending the population to small objects}
\label{sec:truncation}

Outside the adopted population-model support we set $n_C=0$, so hypotheses beyond
the modeled range contribute no likelihood. This introduces a problem at LSST depths: the digest2 model we use ends at $H=25.5$.
At $V\le24$, ordinary main-belt solutions do not reach such faint absolute
magnitudes (Appendix~\ref{app:truncation}), but nearby objects can. Among 499
real NEOCP discovery tracklets (Section~\ref{sec:neocp}), 11 objects (2.2\%)
had their entire admissible region at $H>25.5$. In these cases both
$\Lambda_C$ vanished and the posterior was undefined. All 11 were NEOs, with
median sky motion $6.8^\circ$\,day$^{-1}$ and median $V=20.7$, consistent
with nearby, intrinsically faint objects. digest2 does not encounter the same
hard boundary because its highest-$H$ bin extends to infinity
\citep{keys2019}.

We adopt a similar fix, extending each $(q,e,i)$ cell with one additional bin,
$25.5\le H<30$, assigning it the same density per magnitude as the original
highest bin, $24\le H<25.5$. We continue to set $n_C=0$ for $H\ge30$. This
leaves the original model unchanged below $H=25.5$ while giving the
extrapolated population finite total mass. With this extension, all 11 NEOCP boundary
cases receive a likelihood-based $P(\mathrm{NEO})=1$. Unless stated
otherwise, all results use the extended model.
\medskip

Infrequently, a tracklet can still fall entirely beyond the range of the model. We handle this with an ad-hoc rule: if both
$\Lambda_C$ vanish and at least 95\% of its admissible measure lies at
$H\ge30$, the classifier assigns $P(\mathrm{NEO})=1$ as an explicitly flagged triage
rule rather than a posterior. These assignments are excluded from all
calibration tests. Other cases in which both likelihoods vanish return no
posterior. The rule is motivated by the 11 NEOCP boundary objects above. In
the 800\,000 simulated tracklets of Section~\ref{sec:validation} it fires for
five, all of them NEOs with $q<0.003$~AU, and never for a non-NEO
(Appendix~\ref{app:truncation}).

\subsection{Treatment of measurement uncertainty}
\label{sec:noise}

We treat $A$ and $V$ as exact. This approximation is adequate when the uncertainties in the angular rates
are small compared with the measured motion.

For endpoint position
uncertainties $\sigma$ and an arc length $\Delta t$, the characteristic rate
uncertainty is approximately $\sqrt{2}\sigma/\Delta t$. For high precision
astrometry (such as Pan-STARRS or Rubin) over a 45-minute arc, this is about 0.2\% of the median tracklet motion in our test sample. We test this with a Rubin-like astrometric error model in
Section~\ref{sec:robustness}, and find the approximation adequate for arcs
of at least 15 minutes. Photometric uncertainty, which shifts
the implied $H$ of every hypothesis by the same amount, has not yet been tested
separately but based on end-to-end simulations we don't expect it to play a major role.

We note that measurement uncertainty can be included by marginalizing the class likelihood
over the unknown true observables,
$\int\Lambda_C(A',V')\,p(\mathrm{data}\mid A',V')\,\dd A'\,\dd V'$.
The software interface already supports this form: an attributable may carry
a covariance, and the compiled kernel accepts a weighted set of
attributables. We find this additional marginalization is not necessary in typical operations, which is why we turn it off.

\section{Population model}
\label{sec:population}

\subsection{The input model}
\label{sec:popmodel}

We use the population model maintained by the Minor Planet Center for digest2
\citep{keys2019,veres2023}. It is a histogram in $(q,e,i,H)$ based on the
Synthetic Solar System Model \citep[S3M;][]{grav2011}, with the angular
elements taken to be uniform (Table~\ref{tab:model}). Using the same model in
\pneo and digest2 ensures that comparisons between them reflect the
classification method rather than different population assumptions.

\begin{figure}
\centering
\includegraphics[width=\columnwidth]{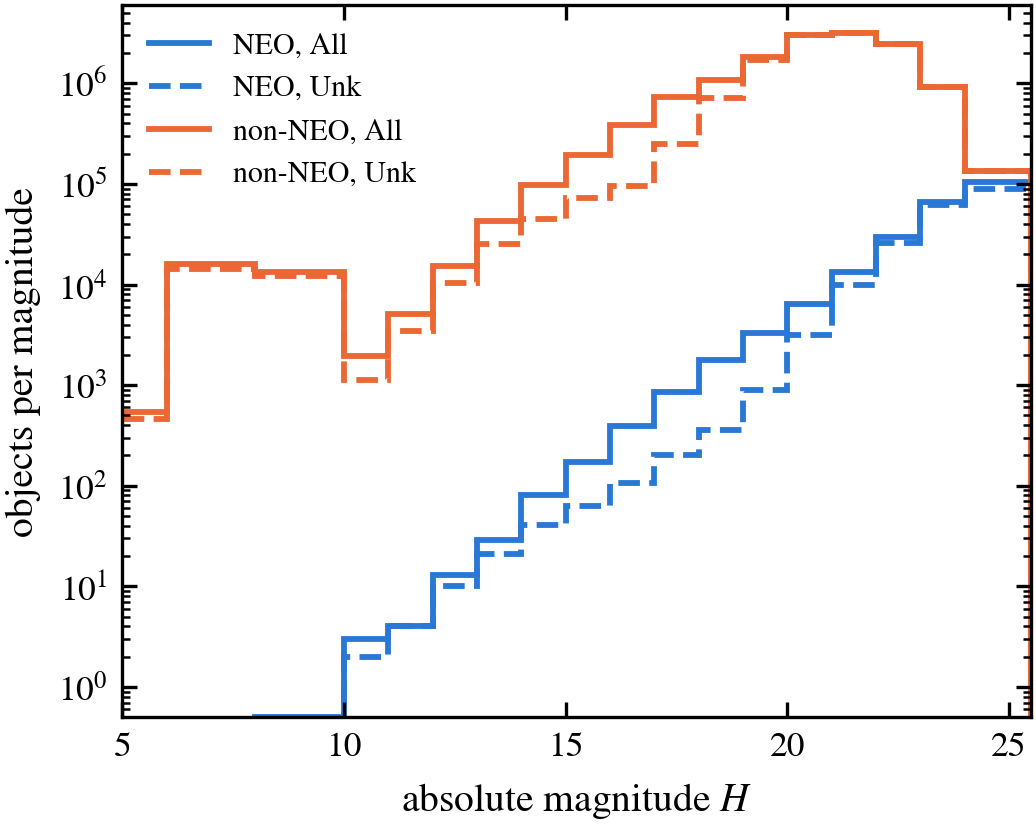}
\caption{Absolute-magnitude distributions of NEOs ($q<1.3$~AU) and
non-NEOs in the digest2 population model. Solid lines show the full modeled
population (All), while dashed lines show the population remaining after
cataloged objects are removed (Unk). The separation between All and Unk
shows the portion of each population that is already known.}
\label{fig:allunk}
\end{figure}

The model file contains two populations. ``All'' is the full modeled
population, while ``Unk'' removes objects already present in the MPC catalog
and therefore represents the population remaining to be discovered.
Figure~\ref{fig:allunk} shows how the two differ in absolute magnitude. The
largest differences occur where a substantial fraction of the modeled
population has already been discovered, while All and Unk become increasingly
similar toward the faint end.

digest2's ``raw'' score uses All, while its ``NoID'' score uses Unk. We use
the corresponding names \pneo All and \pneo Unk. In the simulated validation
and digest2 comparison (Section~\ref{sec:validation}), tracklets are drawn from Unk and scored with
\pneo Unk and digest2 NoID. The real NEOCP candidates
(Section~\ref{sec:neocp}) are also scored with \pneo Unk and digest2 NoID,
because they were undiscovered when posted. Here there is a small mismatch:
Unk is based on the MPC's 2023 catalog, more recent than the 2019--2021
postings, so some objects that were undiscovered at the time have since been
removed from it. We discuss this in Section~\ref{sec:neocp}.

The Rubin tests instead use \pneo All and digest2
raw because those samples include cataloged objects. Strictly, a sample
selected by catalog membership calls for the cataloged population, All
minus Unk, because membership is not a function of the observables and does
not cancel in the posterior. We use All as the closest available model, so
the predicted NEO counts in the known-object samples
(Sections~\ref{sec:lsstknown} and~\ref{sec:dp2}) are an approximate check of
calibration rather than a strict test.

\begin{table}
\centering
\caption{The adopted population model.}
\label{tab:model}
\begin{tabular}{lrl}
\hline
Dimension & Bins & Range \\
\hline
$q$ & 29 & 0--100 AU \\
$e$ & 8  & 0--1.1 \\
$i$ & 11 & 0--180$^\circ$ \\
$H$ & 18 & 0--25.5 (+1 bin to 30, \S\ref{sec:truncation}) \\
\hline
 & \multicolumn{2}{l}{Objects (NEO fraction)} \\
\hline
Total     & \multicolumn{2}{l}{14\,542\,147 \;(1.9\%)} \\
NEO       & \multicolumn{2}{l}{277\,469} \\
Other     & \multicolumn{2}{l}{14\,264\,678} \\
Extended  & \multicolumn{2}{l}{742\,145 NEO, 14\,868\,899 other \;(4.8\%)} \\
\hline
\end{tabular}
\tablecomments{45\,936 cells in total. Counts are after the un-scaling
described in Section~\ref{sec:muk}. ``Extended'' includes the synthetic bin
$25.5\le H<30$.}
\end{table}

\subsection{Recovering absolute counts}
\label{sec:muk}

The digest2 model file does not store the number of objects in each cell
directly. During construction of the model, each cell count is rescaled as
\begin{equation}
\mathrm{cell}=\mathrm{count}\Big/
\sqrt{\Delta q\,\frac{\Delta e}{(1-e_0)+(1-e_1)}\,\Delta i\,\Delta H},
\label{eq:muk}
\end{equation}
where $e_0$ and $e_1$ are the eccentricity-bin edges, $i$ is measured in
degrees, and $1-e$ is set to zero for $e\ge1$. Because $q=a(1-e)$, the
quantity under the square root is, up to a constant factor, the volume $V$
of the cell in $(a,e,i,H)$. The stored value $N/\sqrt{V}$ is therefore
neither the number of objects in the cell, $N$, nor their density, $N/V$,
but the geometric mean of the two. The digest2 source motivates this as an
approximate correction for unequal bin sizes: the admissible region maps
to a two-dimensional surface through the four-dimensional bins, and the
square root of the bin volume stands in for the part of each bin that
surface crosses. The change of measure in \pneo
(Section~\ref{sec:jacobian}) makes this correction exactly.

digest2 works directly with these stored values. \pneo instead requires the
absolute population in each cell, because the relative abundance of the two
classes enters the posterior through the population densities
(Section~\ref{sec:posterior}). We therefore invert Equation~(\ref{eq:muk}) to
recover the original counts.

\subsection{Constructing the  NEO/other split}
\label{sec:split}

The 15 orbit classes used by digest2 do not form a disjoint partition of the
population. Some classes overlap---for example, ``N18'' and ``N22'' are
subsets of ``NEO''---while the belt classes both leave gaps and overlap with
one another. Summing those classes would therefore either omit or double-count
objects.

Instead, we construct two complementary components directly from the model:
NEO $=$ \texttt{NEO}, and other $=$ \texttt{SS}-\texttt{NEO}, where
\texttt{SS} denotes the full Solar System population in the file. Because
$q=1.3$~AU is exactly a bin edge, this split is unambiguous. The NEO component
is zero above the boundary and the other component is zero below it. This is
the complementarity assumed in the definition of
$P(\mathrm{NEO})$ in Section~\ref{sec:posterior}. If the NEO boundary fell
inside a $q$ bin, that bin would contain objects from both classes and the
split would no longer be exact.

\subsection{Known population model limitations}
\label{sec:popcaveats}

The population model has several limitations. First, it is based on the
original S3M population rather than a more recent debiased model. Overall,
S3M predicts about 25\% more objects than are known at $V<20.5$
\citep{wagg2025}, a difference dominated by the main belt; the excess is
larger in the distant populations, where even in the brightest bins the MPC
catalog holds only 14--30\% of the number the model predicts (see also
Figure~1 of \citealt{wagg2025}). Its NEO
component has since been superseded by models such as NEOMOD3 \citep{nesvorny2024}.
Second, the population is represented by a coarse four-dimensional histogram.
Structure within a bin is not modeled, and the flat lowest-$e$ and lowest-$i$
bins interact with the $1/(e\sin i)$ Jacobian factor discussed in
Appendix~\ref{app:jacobian}.

Third, the original magnitude distribution ends at $H=25.5$, with 55.8\% of
the NEO component already contained in the highest-$H$ bin. We address this
cutoff with the extension described in
Section~\ref{sec:truncation}. Finally, the highest eccentricity bin,
$0.9\le e<1.1$, crosses the bound-orbit limit at $e=1$. This does not affect
classification, because the admissible region contains only bound solutions
and the population density is never evaluated at $e\ge1$. It does affect
sampling from the histogram: 1.93\% of NEO draws from All (2.2\% from Unk) and approximately
$2\times10^{-6}$ of other draws are formally unbound. The treatment of these
draws in the validation population is described in
Section~\ref{sec:validation}.

These limitations are properties of the adopted population model rather than of
the classification framework itself. Replacing the population model changes
only the population module, as discussed in Section~\ref{sec:discussion}.
Derivation of an improved population model is left for future work.

\section{Implementation}
\label{sec:implementation}

We provide the classifier of Section~\ref{sec:methods} as an open-source
Python and Rust package, available at \url{https://github.com/lsst-ssp/pneo}
and installable with \texttt{pip install pneo} (version 0.99.0 at the time of
writing). The
software serves two purposes: to validate the method, and to provide a tool
fast enough to score Rubin-era candidate streams in near real time. It
therefore includes two implementations with a common interface.

The reference implementation, in Python, follows the derivation of
Section~\ref{sec:methods} directly. For each range hypothesis it computes
the heliocentric state vector and orbital elements, constructs the full
$6\times6$ Jacobian from analytic derivatives of each step, and evaluates
its determinant numerically. It does not use the closed form of
Equation~(\ref{eq:jacqei}), and so provides an independent check of it.

The production implementation, in Rust, evaluates the most expensive step,
the sum over the grid in Equation~(\ref{eq:lambda}), using the closed-form
Jacobian, and runs in parallel across processor cores. It is called from
Python through PyO3 and parallelized with Rayon. 
Constructing the Jacobian accounts for 88\% of the reference
implementation's run time, and avoiding it is the main source of the
production code's speed. On a single core of an AMD EPYC 7401, the Rust
implementation scores one tracklet on the default $1200\times50$ grid in
about 12~ms, roughly 40 times faster than the reference. Across the 96
threads of a two-processor server, it averages 0.24~ms per tracklet, or
about $1.5\times10^{7}$ tracklets per hour. End to end, scoring the
273\,406 known-object tracklets of the Rubin alert-stream test
(Section~\ref{sec:lsstknown}) with the command-line tool (\texttt{pneo
classify}) takes about
5~minutes per population variant on 90 worker processes of the same
processor, including input parsing and the observer's state, which the
timings above exclude; digest2 takes about 3~minutes.

The two implementations agree to numerical precision. Over 17\,522 grid points drawn from
40 randomized observing geometries, the closed-form and numerical
determinants differ by a median relative amount of $10^{-15}$ and at most
$1.2\times10^{-11}$; over the 20\,000 tracklets of the simulated comparison
(Section~\ref{sec:comparison}), $P(\mathrm{NEO})$ differs by at most
$1.2\times10^{-12}$.

The population model is independent of the classifier: each component
supplies its density, orbital-element coordinates, and domain, so alternative
models can be substituted without other changes.

\section{Validation on Simulated Tracklets}
\label{sec:validation}

We validate \pneo on tracklets simulated from the same population model
that it uses. Section~\ref{sec:selfconsistency} tests whether its
probabilities are calibrated, which checks the calculation itself;
Section~\ref{sec:comparison} compares it with digest2 on a subset of the
same simulated tracklets; and Section~\ref{sec:robustness} tests its
sensitivity to astrometric noise.

\subsection{Calibration and correctness}
\label{sec:selfconsistency}

\subsubsection{Test design}
\label{sec:testdesign}

A correctly implemented classifier should be calibrated when classifying data generated from
the same population model used to compute its posterior. In that case, among
objects assigned $P(\mathrm{NEO})=p$, a fraction $p$ should in fact be NEOs.
We test this directly by drawing objects from the same population model used
by the classifier (Section~\ref{sec:population}) and then classifying the resulting
tracklets.

Because the population model is correct by construction in this experiment,
a systematic failure of calibration would point to the calculation itself,
including the change of measure, admissible-region integration, numerical
quadrature, or implementation. Selection on the measured observables does not
invalidate the test: for example, a magnitude cut multiplies the two class
likelihoods by the same selection function and cancels from the posterior
(Section~\ref{sec:posterior}). This test therefore checks the internal
self-consistency of the method. How well the adopted population model describes the real Solar System (for NEO classification purposes) is tested separately in Section~\ref{sec:neocp}.

\subsubsection{Generating the test tracklets}
\label{sec:generator}

We generate a sample object catalog by drawing $(q,e,i,H)$ from the extended Unk population model and draw the three angular elements uniformly. Each object is placed at a random epoch in 2026, and its attributable and apparent magnitude $V$ are calculated for a geocentric observer. We retain tracklets for objects with $14\leq V<24$, approximating modern surveys.

As the highest eccentricity bin of the population model extends beyond $e=1$, 2.2\% of NEO draws are formally unbound (Section~\ref{sec:popcaveats}). We discard these draws, since the classifier's admissible region contains only bound heliocentric orbits. A rejected object is replaced by a new draw from the full population mixture rather than from the same class, preserving the class proportions of the selected sample.

After the magnitude cut, only 1.3\% of accepted tracklets are NEOs, reflecting the prevalence of non-NEO orbits. We also repeat the test on the All population, with the same results (Section~\ref{sec:gateresults}).

\subsubsection{Calibration tests}
\label{sec:gatestats}

We first test whether the reported $P(\mathrm{NEO})$ values behave as calibrated
probabilities, using two summary statistics and one diagnostic plot.

The first asks whether the classifier predicts the right total number of
NEOs. If $P_j$ is the posterior probability for tracklet $j$ and $y_j=1$ for
an NEO and 0 otherwise, calibration requires the observed count $\sum_j y_j$
to agree with the posterior expectation $\sum_j P_j$. Each tracklet is an
independent yes-or-no outcome, so if the probabilities are correct the
observed count scatters about $\sum_j P_j$ with variance $\sum_j P_j(1-P_j)$.
We measure the difference in units of this expected scatter,
\begin{equation}
Z_{\rm tot} =
\frac{\sum_j (y_j-P_j)}
{\sqrt{\sum_j P_j(1-P_j)}} .
\label{eq:ztot}
\end{equation}
$Z_{\rm tot}$ can be read like a residual in units of $\sigma$:
$|Z_{\rm tot}|\lesssim2$ is consistent with calibration, while
$|Z_{\rm tot}|\gtrsim3$ means that the classifier systematically
over-predicts ($Z_{\rm tot}<0$) or under-predicts ($Z_{\rm tot}>0$) the
number of NEOs. It is the pull of the observed NEO count from its predicted
value, and is closely related to the widely used
goodness-of-fit statistic that bins the sample by predicted probability and
approximates the variance in each bin by that of trials at the bin's mean
probability \citep{hosmer1980}. We use the exact variance instead, because
our $P_j$ span the full range from 0 to 1.

The second statistic measures the size of any miscalibration rather than its
significance. We suppose that the true odds of being an NEO, $P/(1-P)$,
differ from the odds the classifier reports by a constant factor $\hat c$,
and fit $\hat c$ to the labeled sample by maximum likelihood. This is the
logistic recalibration model \citep{cox1958} with its slope fixed at unity.
A value $\hat c=1$ means the probabilities are correct. A value $\hat c>1$
means the classifier understates the probability of being an NEO, and
$\hat c<1$ that it overstates it; a reported $P$ corresponds to a true
probability $\hat cP/(1-P+\hat cP)$. For example, $\hat c=1.1$ changes a
reported $P=0.50$ to 0.52, and $P=0.10$ to 0.11. We quote 95\% intervals on
$\hat c$ from the Gaussian approximation to its likelihood. Tracklets with
$P(\mathrm{NEO})$ exactly 0 or 1 do not constrain $\hat c$, because no finite
rescaling of the odds changes them, and we leave them out of the fit.

Finally, a reliability diagram provides a complementary check across the
full range of $P(\mathrm{NEO})$ by comparing the observed NEO fraction with
the mean predicted probability in bins of width 0.1
(Figure~\ref{fig:reliability_unk}). It is the counterpart, for a yes-or-no
outcome, of the calibration tests applied to probabilistic photometric
redshifts \citep[e.g.,][]{wittman2016,disanto2018}.

The statistical precision of the results to be discussed in the next section
is limited mainly by the number of NEOs in the sample.
A run of $2\times10^5$ Unk tracklets contains about 2\,600 NEOs and
constrains $\hat c$ to about $\pm14\%$; combining four independent runs
improves this to about $\pm7\%$. Interpreting the presented numbers
should take this into account.

\subsubsection{Results}
\label{sec:gateresults}

\begin{table}
\centering
\caption{Self-consistency results for the extended population model.}
\label{tab:gate}
\setlength{\tabcolsep}{4pt}
\begin{tabular}{lccc}
\hline
Run & NEOs obs./exp. & $Z_{\rm tot}$ & $\hat c$ [95\%] \\
\hline
Unk 1       & 2591 / 2599.2 & $-0.55$ & 0.96 [0.84, 1.10] \\
Unk 2       & 2675 / 2654.6 & $+1.40$ & 1.10 [0.96, 1.25] \\
Unk 3       & 2642 / 2650.3 & $-0.57$ & 0.96 [0.84, 1.10] \\
Unk 4$^a$   & 2642 / 2638.6 & $+0.23$ & 1.02 [0.89, 1.16] \\
Unk pooled  & 10\,550 / 10\,542.6 & $+0.25$ & 1.01 [0.94, 1.08] \\
\hline
All A + B   & 1719 / 1743.8 & $-1.42$ & 0.92 [0.82, 1.03] \\
\hline
\end{tabular}
\tablecomments{Each Unk run simulates $2\times10^5$ tracklets with a
different random seed; All A + B is two runs of $10^5$. $^a$Epochs in 2027;
the other runs use 2026. Excluded are tracklets beyond the 100~AU
integration limit (52 Unk, 5 All; Section~\ref{sec:ar}), tracklets with no
posterior (37 Unk, 3 All; all sungrazing NEOs with $q<0.004$~AU), and five
Unk NEOs assigned $P(\mathrm{NEO})=1$ by the triage policy
(Section~\ref{sec:truncation}).}
\end{table}

\begin{figure}
\centering
\includegraphics[width=\columnwidth]{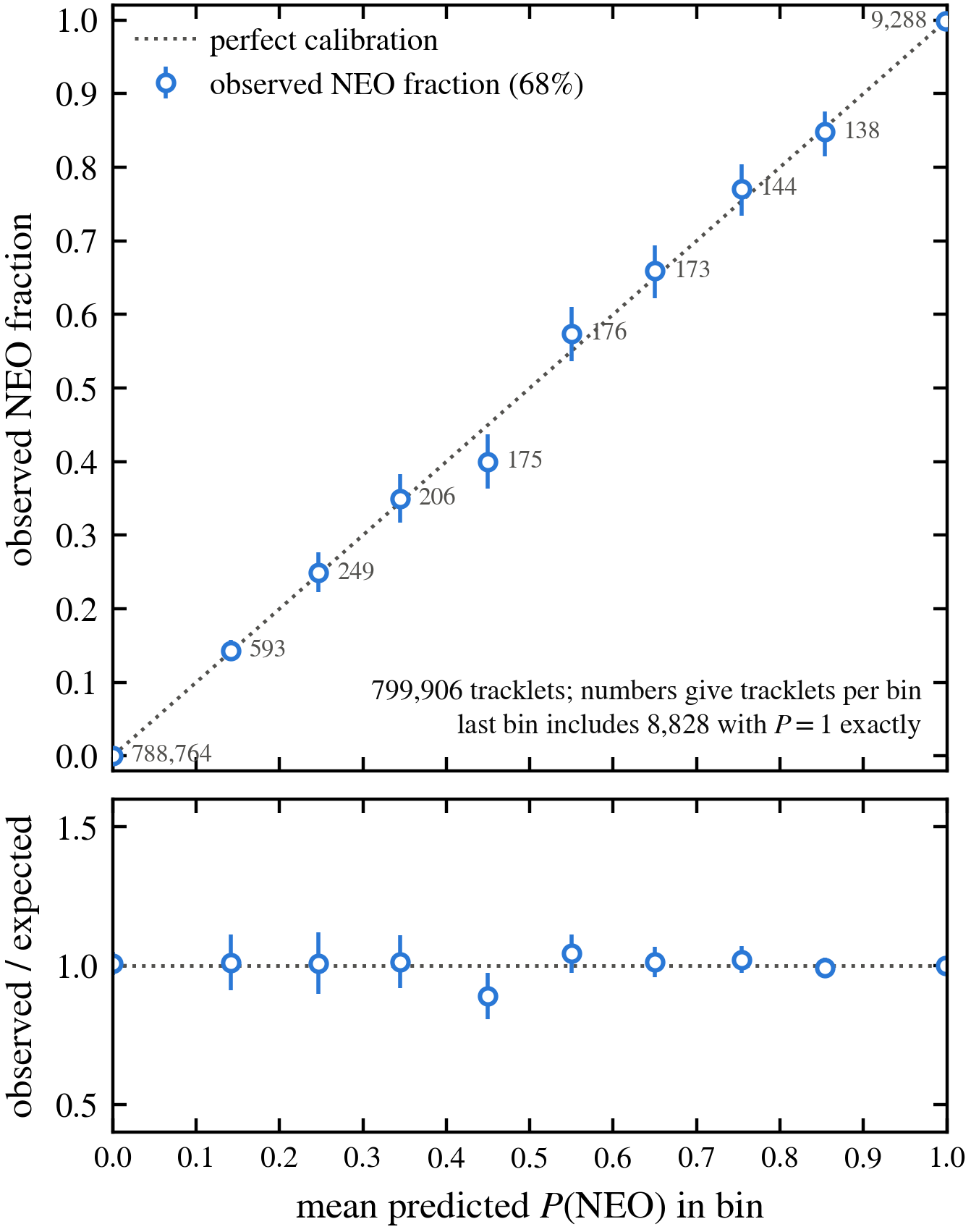}
\caption{Reliability of $P(\mathrm{NEO})$ for tracklets drawn from and
scored with the undiscovered (``Unk'') population, combining four
independent samples for a total of 799\,906 tracklets. Tracklets are grouped
into bins of width 0.1 in predicted $P(\mathrm{NEO})$. Top: observed NEO
fraction versus mean predicted probability in each bin, with 68\% intervals;
the dotted line shows perfect calibration. Numbers beside the points give the
number of tracklets in each bin. The final bin contains 9\,288 tracklets,
including 8\,828 with $P(\mathrm{NEO})=1$ exactly. Bottom: ratio of the
observed to expected NEO count in each bin, where the expected count is
$\sum P(\mathrm{NEO})$; the dotted line marks unity and the error bars show
the 68\% scatter expected for a calibrated classifier. Overall, 10\,550 NEOs
are observed compared with 10\,542.6 expected, with
$\hat c=1.01$ [0.94, 1.08].}
\label{fig:reliability_unk}
\end{figure}

The classifier is calibrated on its own population model. Pooling the four
Unk runs, 10\,550 NEOs are observed against 10\,542.6 predicted
($Z_{\rm tot}=0.25$), with an odds scale $\hat c=1.01$ [0.94, 1.08]; each
individual run, and the All runs, are likewise consistent with calibration
(Table~\ref{tab:gate}). The agreement holds across the full range of
probabilities (Figure~\ref{fig:reliability_unk}). At the extreme, all
8\,828 tracklets assigned exactly $P(\mathrm{NEO})=1$, because no non-NEO
solution has population support, are NEOs.

The inference of the unobserved distance is also accurate. For every
simulated tracklet within the integration domain, the true
$(\rho,\dot\rho)$ lies inside the admissible region, and for the median
tracklet 99.98\% of the ranging posterior lies within a factor of two of
the true distance.

\subsection{Comparison with digest2}
\label{sec:comparison}

\label{sec:premise}%
We compare \pneo with digest2 (version 2.8.2;
\citealt{keys2019,veres2023}) on 19\,999 Unk tracklets, 287 of them NEOs,
generated as in Section~\ref{sec:generator}.\footnote{Each tracklet is
supplied to digest2 as two detections 45~minutes apart, in ADES format at
full precision, from a geocentric site. Both methods use the population
through $H=30$; about 2\% of tracklets place more than 1\% of their
admissible measure beyond it, where digest2's last $H$ bin continues
indefinitely. One further tracklet could not be written for digest2 and is
excluded from both.} \pneo uses the Unk population and digest2 its NoID
score, which is computed from the same population, so the comparison
isolates the classification method.

The digest2 score is not a probability. If read as one (NoID/100), it predicts
1349 NEOs in this sample, 4.7 times the 287 present; the \pneo posteriors
predict 284.0, and all 237 tracklets with $P(\mathrm{NEO})=1$ are NEOs.
This is not a failure of digest2, whose score is designed for ranking, but
it means the score cannot be interpreted directly as an NEO probability.
The \pneo output, in contrast, is a calibrated probability, and so gives the
expected return of a follow-up observation directly. This matters for
multi-purpose facilities that must weigh a planetary-defense target against
other time-critical programs, such as gravitational-wave counterparts or
other transients: a probability can be combined with the value and cost of
each observation on a common scale, whereas a ranking score can only order
NEO candidates among themselves.

\begin{figure}[t]
\centering
\includegraphics[width=\columnwidth]{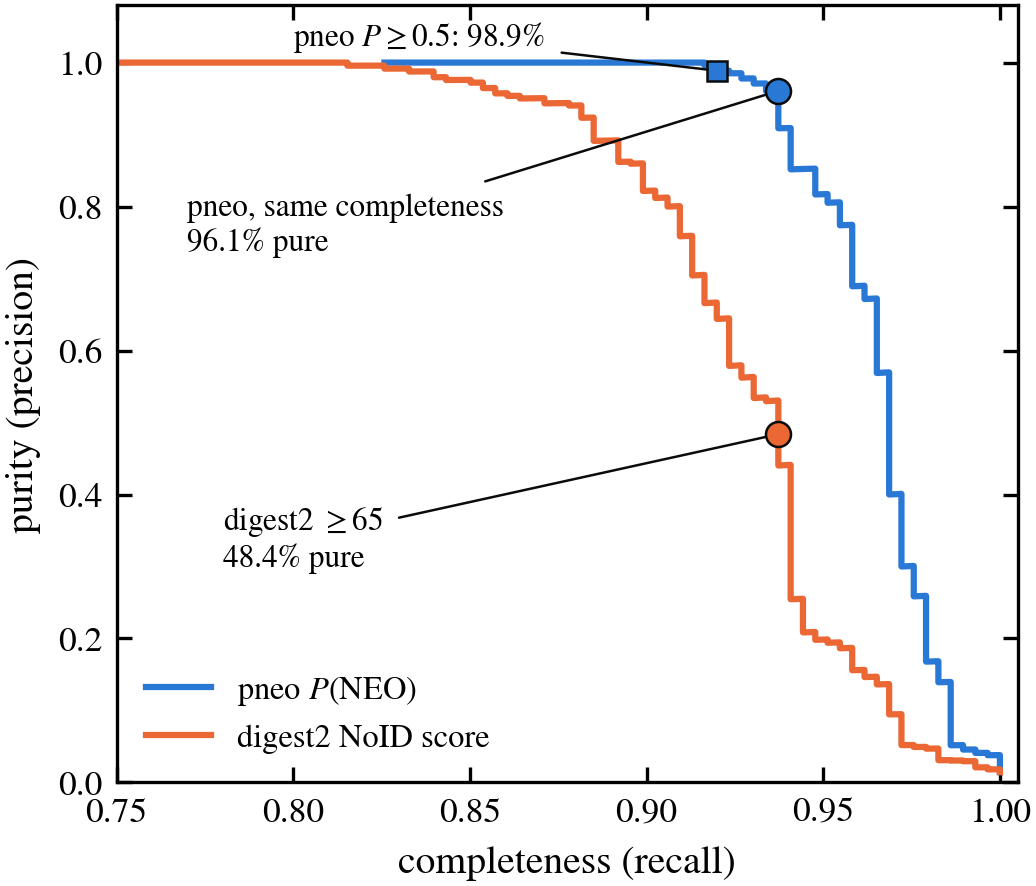}
\caption{Purity (the fraction of selected tracklets that are NEOs) against
completeness for \pneo and digest2 (NoID score) on the 19\,999 simulated
Unk tracklets of Section~\ref{sec:premise} (287 NEOs). Circles mark the digest2 NEOCP threshold of 65, which is
48.4\% pure at 93.7\% completeness, and \pneo at the same completeness, which
is 96.1\% pure. The square marks \pneo $P(\mathrm{NEO})\geq0.5$, 98.9\% pure
at 92.0\% completeness.}
\label{fig:d2pr}
\end{figure}

The second consideration for follow-up is how many NEOs are recovered at a
given contamination (the purity), minimizing time wasted on observations of
non-critical objects. Non-NEOs outnumber NEOs by 69 to 1 in this sample, so a few
highly ranked non-NEOs can strongly reduce purity. At the NEOCP posting threshold, digest2 selects 269 NEOs and 287 non-NEOs;
at the same completeness, \pneo selects the same number of NEOs, 267 of
them in common, but only 11 non-NEOs (purity 48\% and 96\%, respectively;
Table~\ref{tab:ops}). The advantage is not specific to this threshold: at
every completeness, \pneo is at least as pure as digest2
(Figure~\ref{fig:d2pr}).

\begin{table}
\centering
\caption{Operating points on the shared sample of 287 NEOs.}
\label{tab:ops}
\setlength{\tabcolsep}{4pt}
\begin{tabular}{lrrcc}
\hline
Cut & NEO & Non-NEO & Compl. & Contam. \\
\hline
digest2 NoID $\geq65$   & 269 & 287 & 93.7\% & 51.6\% \\
\pneo $P\geq0.30$$^a$    & 269 & 11  & 93.7\% & 3.9\% \\
\pneo $P\geq0.5$         & 264 & 3   & 92.0\% & 1.1\% \\
\hline
\end{tabular}
\tablecomments{NEO and Non-NEO give the numbers of selected tracklets of
each class; Compl.: completeness; Contam.: contamination. $^a$Threshold
chosen to match the 93.7\% completeness of the digest2 NoID $\geq65$ cut.}
\end{table}

\begin{figure*}[t]
\centering
\includegraphics[width=0.88\textwidth]
{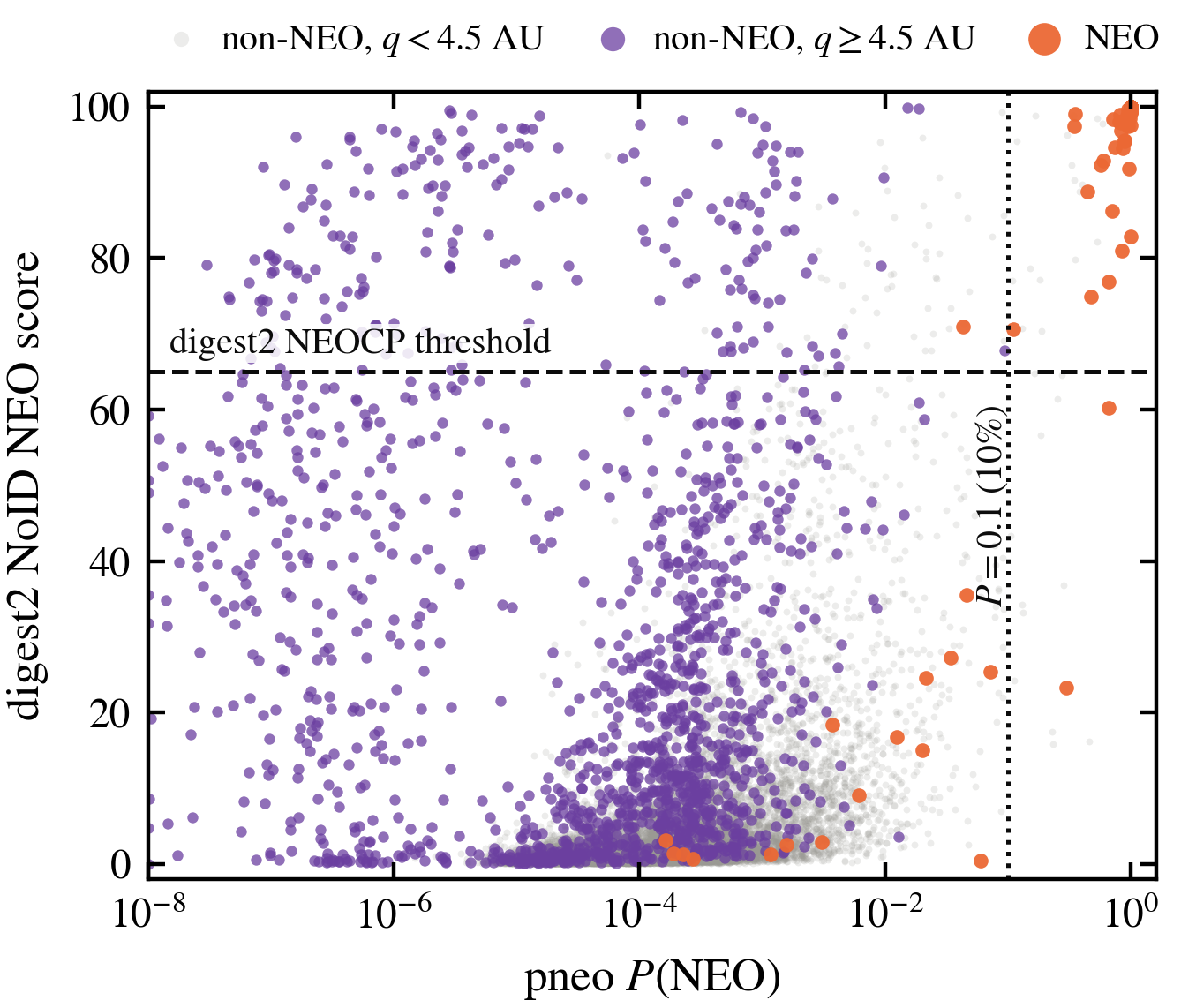}

\caption{digest2 NoID score versus \pneo $P(\mathrm{NEO})$ for the 19\,999
simulated Unk tracklets of Section~\ref{sec:premise} (287 NEOs). Points are colored
by true class, with non-NEOs separated at $q=4.5$~AU. The horizontal dashed
line marks the digest2 NEOCP threshold of 65, and the vertical dotted line
marks $P(\mathrm{NEO})=0.1$. Probabilities below $10^{-8}$ are plotted
at $10^{-8}$.}
\label{fig:d2scatter}
\end{figure*}

\label{sec:disagreement}%
The difference is almost entirely due to distant objects
(Figure~\ref{fig:d2scatter}). Of digest2's 287 false positives, 213 have
$q\ge4.5$~AU, at heliocentric distances of 13--89~AU, and a further 58 lie
in the main belt ($1.67\le q<4.5$~AU). \pneo assigns all 213 distant objects
$P(\mathrm{NEO})<0.1$, most below $10^{-4}$, and selects none with
$q\ge4.5$~AU. The slow tracklet of a distant object admits nearby, NEO-like
solutions as well as its true distant one. digest2 counts the full
population of every NEO bin those nearby solutions reach, whereas in \pneo
the $\rho^4$ factor (Section~\ref{sec:jacobian}) gives the distant
solutions far more phase-space volume.

The remaining \pneo false positives
lie close to the NEO boundary: ten of the 11 are Mars-crossers with
$1.3\le q<1.67$~AU, and one is in the main belt. Their follow-up, even though
they're not NEOs, would still result in recovery of objects more interesting
than a typical distant asteroid or TNO.

The digest2 excess is not
caused by its assumed astrometric error: adopting the $0.1''$ it assigns to
Rubin instead of its default $1''$ increases its false positives on the All
sample from 221 to 276, 73\% of them distant.
Section~\ref{sec:neocp} repeats the comparison on real discovery tracklets.

\subsection{Robustness to astrometric noise}
\label{sec:robustness}

The classifier treats the attributable as exact (Section~\ref{sec:noise}).
To test this at Rubin's astrometric precision, we perturb both detections
of every tracklet in the All self-consistency sample (20\,000 tracklets,
170 NEOs) with the Rubin astrometric error model of the Sorcha survey
simulator \citep{merritt2025}, which gives per-coordinate uncertainties of
10--85~mas over $V=18$--24 (median 35~mas).\footnote{The uncertainty in each
coordinate is $\sigma=\sqrt{\sigma_{\rm sys}^2+(0.6\,\theta/\mathrm{SNR})^2}$,
with a systematic floor $\sigma_{\rm sys}=10$~mas, seeing $\theta=0.7''$,
and the signal-to-noise ratio from the LSST photometric model
\citep{ivezic2019} with $m_5=24$ and trailing losses \citep{veres2017}. We
apply this prescription to our simulated detections rather than generating
the tracklets with Sorcha.} We then recompute the angular rates and rescore
the tracklets.

For 45-minute arcs the effect is negligible: 99\% of tracklets change
$P(\mathrm{NEO})$ by less than $7\times10^{-4}$. Shorter arcs turn the same
positional error into a larger rate error, scaling as $1/\Delta t$
(Table~\ref{tab:arclength}). Down to 15~minutes, the predicted number of
NEOs and the contamination stay close to the noiseless values; at
10~minutes contamination rises to 12.5\%, and at 5~minutes the classifier
is no longer calibrated ($Z_{\rm tot}=-4.5$). The cost of short arcs is
contamination rather than lost NEOs. Most of the added false positives are
slow trans-Neptunian objects, whose small apparent motion is changed by
tens of percent by the same astrometric error, enough to make nearby
NEO-like solutions plausible.

\begin{table}
\centering
\caption{Scores with Rubin-like astrometric noise as the arc shortens.}
\label{tab:arclength}
\begin{tabular}{rcccc}
\hline
Arc & Rate error$^a$ & $\sum P$ & $Z_{\rm tot}$ & Contam.$^b$ \\
(min) & & (170 NEOs) & & \\
\hline
none$^c$ & --        & 166.1 & $+0.71$ & 6.3\% \\
60       & 0.14\%    & 166.2 & $+0.69$ & 6.4\% \\
30       & 0.28\%    & 166.3 & $+0.67$ & 6.4\% \\
15       & 0.57\%    & 170.2 & $-0.04$ & 8.7\% \\
10       & 0.85\%    & 176.0 & $-1.06$ & 12.5\% \\
5        & 1.70\%    & 197.9 & $-4.49$ & 25.6\% \\
\hline
\end{tabular}
\tablecomments{$^a$Median rate error as a fraction of the tracklet's own
motion. $^b$Contamination at the fixed cut $P\ge0.214$, the cut that matches digest2's completeness on the All sample; completeness stays at 86.5--87.1\% in every row.
$^c$Noiseless scores.}
\end{table}

These arc lengths overlap those expected for Rubin. In the first year of
the v5.3.7 baseline cadence simulation \citep{bianco2022,PSTN-056}, the
median span of same-night visits to a field is 23~minutes; 70\% of
tracklets span at least 15~minutes and 15\% less than 10~minutes. Treating
the attributable as exact is therefore adequate for most Rubin tracklets;
for shorter arcs, the measurement uncertainty should be marginalized
(Section~\ref{sec:noise}).

\section{Validation on Real Data}
\label{sec:realdata}

We next test the classifier on real tracklets. We start with the NEO Confirmation Page sample (Section~\ref{sec:neocp}), consisting of real discovery tracklets but already filtered by digest2. To test the classifier without this
filtering, we next run on detections from the Vera C.\ Rubin Observatory alert stream (Section~\ref{sec:alerts}) and from Rubin Data Preview~2
(Section~\ref{sec:dp2}). Section~\ref{sec:rubinsummary} summarizes the
results.

\subsection{The NEO Confirmation Page}
\label{sec:neocp}

We test the classifier on real discovery tracklets from the NEO Confirmation Page
(NEOCP). The MPC's archive of past NEOCP postings\footnote{\url{https://www.minorplanetcenter.net/iau/NEO/elsewhere.csv}}
lists 17\,901 postings between 2019 March 21 and 2021 December 21. We keep
the 13\,471 that later received a designation, are not comets, and were
discovered from a ground-based station, and draw 500 of them at random. For
each object we retrieve the astrometry reported on the discovery night from
the MPC observations service\footnote{\url{https://data.minorplanetcenter.net/api/get-obs}},
using the detections from the discovering station only; 499 objects yield a
usable tracklet, with a median of four detections spanning 35~minutes.
Objects are labeled using their current orbits, with $q<1.3$~AU defining an
NEO, taking $q$ from the MPC's list of near-Earth asteroids and checking it
against the JPL Small-Body Database; the sample contains 305 NEOs and 194
non-NEOs.

This is not an unbiased sample of newly detected moving objects. Every object
reached the NEOCP after receiving a sufficiently high digest2 score at the MPC
and passing submitter review, so the sample is already strongly selected toward
NEO-like tracklets. The requirement of a designation adds a second
selection. About 15\% of postings were never designated, either because
follow-up failed or because the detection was spurious. Recovery depends on
the object's true orbit as well as on its tracklet: nearby, fast-moving NEOs
are the easiest to lose. Unlike the digest2 posting cut, this selection need
not cancel in the posterior, and it may slightly lower the observed NEO
fraction relative to the predicted one. Its effect on calibration is
limited, however. The objects most easily lost, nearby and fast-moving, are
also those to which the classifier already assigns $P(\mathrm{NEO})\approx1$,
so losing them removes predicted and observed NEOs alike; only losses that
differ by class among tracklets with intermediate probabilities can bias the
comparison. Part of the undesignated 15\% are also spurious detections, which
belong to neither class. Consistent with this, the predicted and observed NEO
counts below agree to about 2\%.

\paragraph{Calibration.}
As in Section~\ref{sec:validation}, we score the
sample with \pneo Unk and compare with digest2's NoID score. Across the 499
tracklets, \pneo predicts 312.0 NEOs and 305 are observed.

\begin{table}
\centering
\caption{Operating points on the NEOCP sample of 305 NEOs (\pneo Unk,
digest2 NoID).}
\label{tab:neocp}
\setlength{\tabcolsep}{4pt}
\begin{tabular}{lrrcc}
\hline
Cut & NEO & Non-NEO & Compl. & Contam. \\
\hline
digest2 NoID $\geq65$   & 300 & 140 & 98.4\% & 31.8\% \\
\pneo $P\geq0.030$$^a$   & 300 & 85  & 98.4\% & 22.1\% \\
\pneo $P\geq0.5$         & 286 & 22  & 93.8\% & 7.1\% \\
\hline
\end{tabular}
\tablecomments{NEO and Non-NEO give the numbers of selected tracklets of
each class. digest2 is re-run locally on the reconstructed discovery
tracklets, so 11.8\% of the sample falls below 65. Completeness is relative
to the NEOs in the NEOCP sample. $^a$Threshold chosen to match the
completeness of the digest2 $\geq65$ cut.}
\end{table}

\paragraph{Ranking.}
Because the sample was preselected by digest2, it lacks the tracklets that
digest2 scores low, and the ranking comparison is not unbiased. At the
completeness reached by digest2 $\ge65$ cut, \pneo selects
55 fewer tracklets (Table~\ref{tab:neocp}). Its false positives also remain
close to the NEO boundary: all 22 non-NEOs with
$P(\mathrm{NEO})\ge0.5$ have $q<2.3$~AU, and 17 have
$q<1.55$~AU. The digest2 false positives in this real sample are likewise
mostly nearby in orbital-element space, with 91 of 140 having
$q<1.67$~AU and only four having $q\ge4.5$~AU.
This differs from the synthetic comparison, where most of digest2's false
positives were distant objects. The difference is not caused by digest2's
preselection, since distant objects in the simulation score above the
posting threshold; instead, as we show next,
the bright distant objects that dominate the simulated false positives
are already cataloged in reality and are removed before submission.

\begin{figure}
\centering
\includegraphics[width=\columnwidth]{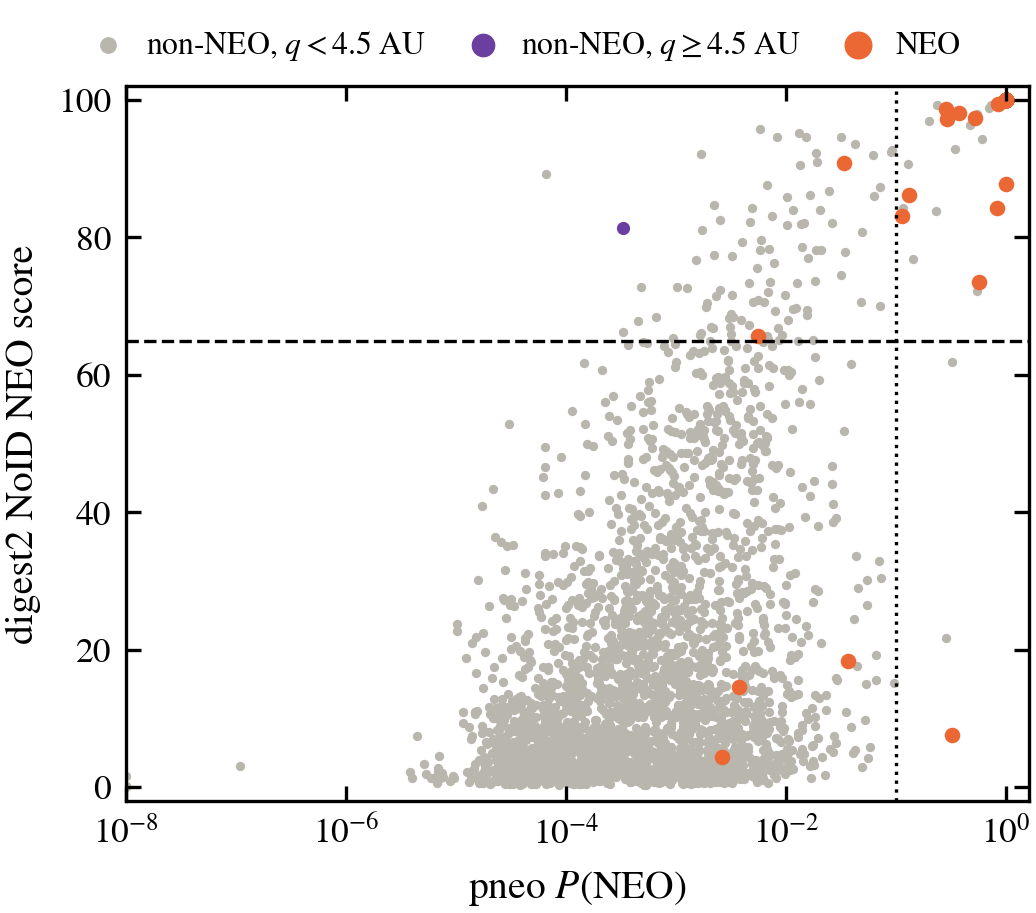}
\caption{digest2 NoID score against \pneo $P(\mathrm{NEO})$ for a simulated
Unk sample limited to $V<21.2$, after removing distant objects ($q\geq4.5$~AU)
bright enough to be cataloged (brighter than the peak of the known
population's $H$ distribution in the same perihelion range). The dashed line
marks the digest2 NEOCP threshold of 65 and the dotted line
$P(\mathrm{NEO})=0.1$. At digest2 $\ge65$ the sample has 33.9\% contamination
at 98.1\% completeness, against 31.8\% at 98.4\% on the real NEOCP sample.}
\label{fig:neocpbright}
\end{figure}

\paragraph{Why the NEOCP contains few distant false positives.}
In the simulated comparison, 213 of digest2's 287 false positives have
$q\geq4.5$~AU, and its contamination at the NEOCP threshold is 51.6\%
against 3.9\% for \pneo at the same completeness (Table~\ref{tab:ops}).
Two things separate the NEOCP sample from the simulated one: the surveys
that fed it were shallower, and their observers remove objects that are
already cataloged before submitting candidates. To test whether these
account for the difference, we drew a second simulated sample from the Unk
population limited to $V<21.2$, 0.1~mag fainter than the median of the NEOCP
sample, and scored it with both classifiers (3\,498 tracklets, 215 NEOs). The
brightness limit alone does not reproduce the NEOCP: at digest2 $\ge65$ the
contamination is still 54.2\%, with 143 false positives at $q\geq4.5$~AU
(Table~\ref{tab:neocpbright}). We then removed the distant objects
($q\geq4.5$~AU) that are bright enough to be cataloged, taken as those with
$H$ brighter than the peak of the $H$ distribution of known objects in the
same range in the MPC orbit catalog ($H=14.75$ for $4.5\le q<10$~AU and
$H=7.25$ for $q\geq10$~AU). This removes all but one of the distant false
positives and gives 33.9\% contamination at 98.1\% completeness, close to
the 31.8\% at 98.4\% measured on the NEOCP
(Figure~\ref{fig:neocpbright}; Table~\ref{tab:neocpbright}). \pneo's
contamination at the same completeness, 17.3\%, is likewise close to the
22.1\% on the NEOCP. 

\begin{table}
\centering
\caption{The NEOCP operating point reproduced in simulation (digest2 NoID
$\ge65$; \pneo Unk).}
\label{tab:neocpbright}
\setlength{\tabcolsep}{4pt}
\begin{tabular}{lccc}
\hline
Sample & Compl. & Contam. & FP \\
\hline
Simulated, $V<24$              & 93.7\% & 51.6\% & 213 \\
Simulated, $V<21.2$            & 98.1\% & 54.2\% & 143 \\
\quad known distant removed    & 98.1\% & 33.9\% & 1 \\
NEOCP (real)                   & 98.4\% & 31.8\% & 4 \\
\hline
\end{tabular}
\tablecomments{Completeness and contamination of the digest2 $\ge65$
selection; FP: its false positives with $q\geq4.5$~AU. \pneo at the same completeness has 3.9\%, 17.6\%,
17.3\% and 22.1\% contamination in the four rows.}
\end{table}

The NEOCP is therefore protected today by the completeness of the MPC
catalog: the main-belt and distant objects bright enough to be detected by
the surveys that feed it are already known, and are removed before
submission. This protection does not extend to Rubin's depth. LSST will survey to
$V\approx24$ and is expected to discover 4--9 times more objects than are
presently known in each population, including the main belt, the Jupiter
Trojans, and the trans-Neptunian objects \citep{kurlander2025}.
\citet{wagg2025} show that, under the current digest2 criterion, most Rubin
NEOCP candidates would be such undiscovered main-belt asteroids. The
$V<24$ comparison of Section~\ref{sec:comparison} reproduces this regime,
and -- combined with this analysis -- shows why digest2's selection degrades in it: without the catalog to
remove them, distant and main-belt objects dominate its false positives.

\subsection{The Rubin alert stream}
\label{sec:alerts}

For every new image, Rubin subtracts a template of the same field and
issues an alert for each source detected in the difference image, within
minutes of the observation \citep{LDM-612}. The alerts are distributed to
the community through brokers such as Fink \citep{moller2021} and ALeRCE
\citep{forster2021} from which anyone can download them. Known Solar System objects are tagged within the alert itself; unknown ones are (by
definition) not. The alert stream is
therefore the earliest data in which Rubin's NEO candidates appear, and the
natural place to apply a classifier.

\subsubsection{Data and processing}
\label{sec:lsstdata}
\label{sec:lsstlink}
\label{sec:lsstlabels}

We use alerts from eleven nights in 2026 July and October, obtained through
the Fink broker \citep{moller2021}, with standard quality cuts on flux,
image-artifact flags, and real/bogus reliability
(Appendix~\ref{app:rubindata}). We link detections blindly into
single-night tracklets with HelioLinC's \texttt{make\_tracklets}
\citep{heinze2022AAS}, attaching object identities only after linking. Our linking reproduces Rubin's own associations: among known objects, 99.5\%
of object-nights with at least two detections, and 111 of 112 for NEOs, are
recovered as \emph{pure} tracklets, in which all detections belong to the
same object (tracklets combining detections of different objects we call
\emph{mixed}). This tests only the linking; classification is tested below.
We retain arcs of at least 15~minutes
(Section~\ref{sec:robustness}) and remove tracklets that move less than
$0.5''$ in total, which are repeated detections of stationary sources.
Known-object tracklets are labeled from their MPC designation and the MPC
orbit catalog \citep{mpcorb}; unassociated tracklets are labeled by
propagating every cataloged orbit to their detections. Every tracklet is
scored with \pneo All and Unk and with digest2 raw and NoID.

\paragraph{Removing static sources.}
\label{sec:lsstcut}
Static sources are the main contaminant among unassociated tracklets: a
variable star detected several times in one night, or an unrelated detection
near such a source, can form a plausible-looking tracklet that neither
classifier can distinguish from a slow mover. We therefore remove
unassociated detections whose diaObject (Rubin's record of an individual
source seen in difference images; \citealt{LSE-163}) has more than two
detections. This removes them without losing asteroids. Because an asteroid moves between visits, Rubin normally creates a new
diaObject for each of its detections: among 6\,793 unassociated detections
within $2''$ of a cataloged asteroid, only four (0.06\%) belong to
diaObjects with more than two detections, compared with 25.2\% of all
unassociated detections. The cut does not change the labeled results.

\subsubsection{Known objects}
\label{sec:lsstknown}

After the 15-minute cut, the known sample contains 273\,406 tracklets, of
which 64\,938 are pure (81 of them NEOs) and 208\,468 mixed. We evaluate
the classifiers on the pure tracklets (Table~\ref{tab:lsstknown}).
To compare them fairly, we fix the number of NEOs each one selects and
count how many non-NEOs it selects along with them. At its NEOCP posting
threshold of 65, digest2 selects 57 of the 81 NEOs; \pneo selects almost
the same number, 56, at $P\geq0.1$. To do so, digest2 also selects 69
non-NEOs, 19 of them distant objects with $q\geq4.5$~AU, whereas \pneo
selects 17 non-NEOs and no distant objects.

\begin{table}
\centering
\caption{Known objects in Rubin alerts for pure tracklets with arcs
$\geq15$~min. ``Predicted'' is $\sum P$ for \pneo All. Flagged counts are
NEO / non-NEO / non-NEO with $q\geq4.5$~AU.}
\label{tab:lsstknown}
\begin{tabular}{lcc}
\hline
 & 2+ detections & 3+ detections \\
\hline
Tracklets / NEOs & 64\,938 / 81 & 10\,266 / 12 \\
\pneo All predicted & 68.9 & 13.0 \\
digest2 raw $\geq65$ & 57 / 69 / 19 & 12 / 7 / 1 \\
\pneo All $P\geq0.1$ & 56 / 17 / 0 & 12 / 2 / 0 \\
\pneo All $P\geq0.5$ & 47 / 4 / 0 & 8 / 0 / 0 \\
\hline
\end{tabular}
\end{table}

\pneo also predicts the number of NEOs reasonably well: 68.9 against 81
observed. The shortfall is concentrated in distant Amors with slow,
main-belt-like motion: of the 21 NEO tracklets with $P<0.01$, 13 are Amors,
with semimajor axes up to 3.1~AU. Because these tracklets are selected by
catalog membership, a selection the posterior does not cancel
(Section~\ref{sec:population}), the known population is enriched in such
NEOs relative to the model, and a deficit of this kind is expected.

On real Rubin detections of known objects, \pneo therefore recovers as many
NEOs as digest2 while selecting about four times fewer non-NEOs, and none
of the distant objects that dominate digest2's false positives.

The mixed tracklets, pairs of detections belonging to different objects, are a
separate problem that no tracklet-based classifier can solve. Nearly
all have only two detections: only 12 of 208\,468 have three or more, so
requiring three or more detections serves as a very effective filter to remove them.

\subsubsection{Unassociated detections with labels}
\label{sec:lsstunknown}

The unassociated alerts yield 171\,859 moving tracklets. Most are objects
with no known orbit, but 3\,345 (22 of them NEOs) turn out to match
cataloged objects that Rubin did not associate at the time of the alert.
These provide a labeled test close to discovery conditions: the
classifiers see them exactly as they would see a new object. Among them,
208 tracklets of 165 objects (12 of them NEO tracklets) belong to objects
that were designated only after the alert, and so were genuinely new at
the time.

\pneo predicts 22.8 NEOs against 22 observed (Table~\ref{tab:lsstunk}).
Compared at equal NEO recovery, as before, \pneo selects 21 NEOs at
$P\geq0.1$ and digest2 20 at its threshold of 65; digest2 also selects 29
non-NEOs, 17 of them distant objects with $q\geq4.5$~AU, whereas \pneo
selects 14 non-NEOs and no distant objects. All 17 of digest2's distant
false positives receive $P<0.01$. For the 208 tracklets of newly designated
objects, \pneo predicts 11.3 NEOs against 12 observed. In conditions close to discovery, \pneo therefore remains calibrated and
again selects half as many non-NEOs as digest2, none of them distant.

\begin{figure*}
\centering
\includegraphics[width=\textwidth]{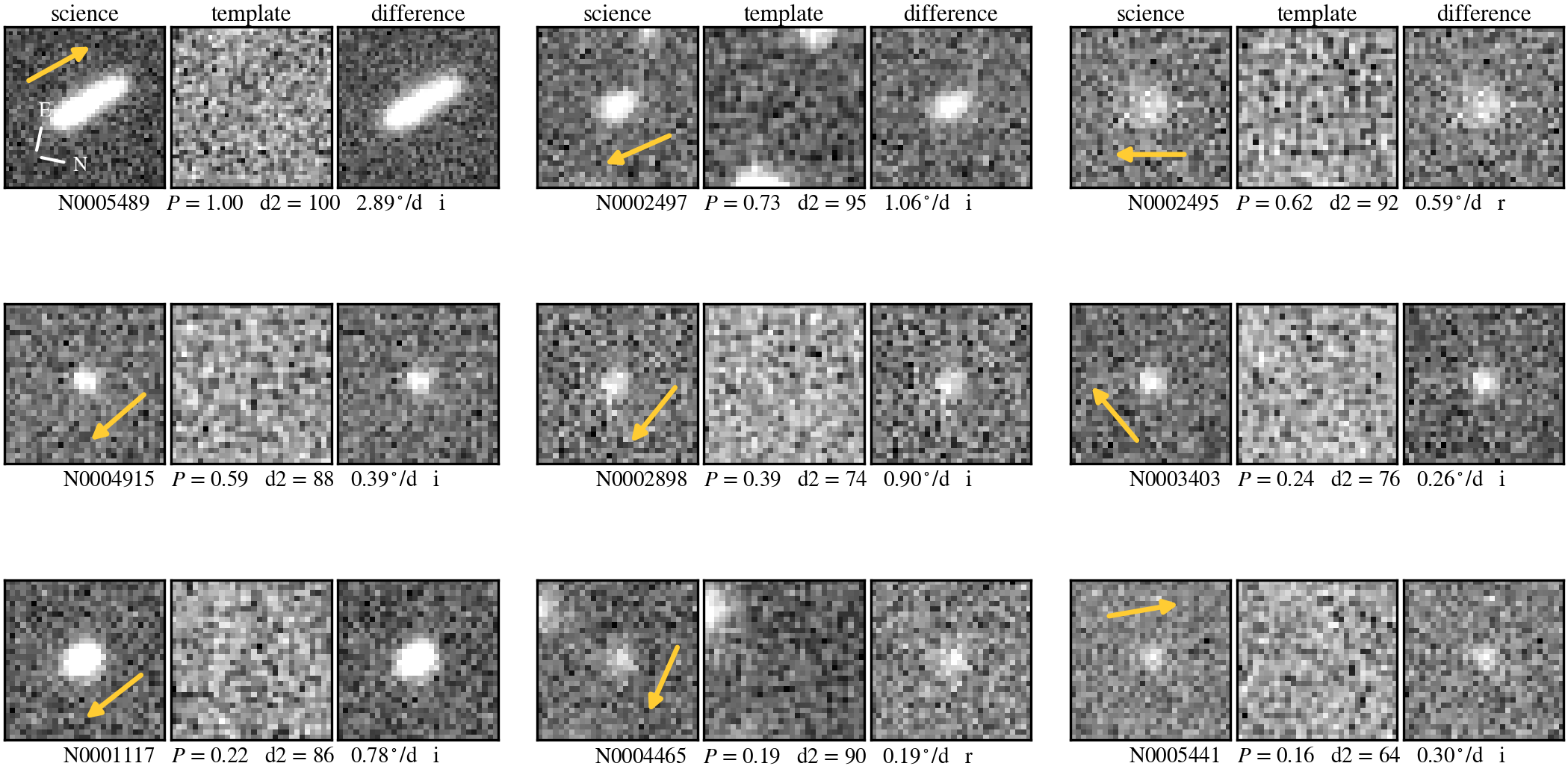}
\caption{Rubin alert cutouts of nine unassociated tracklets with no matching
known orbit that image inspection confirms as real moving objects, at each
tracklet's highest signal-to-noise detection: science,
template and difference images from the ALeRCE broker \citep{forster2021}. The source appears in
the science and difference images but not in the template. Arrows show the
direction of motion over the tracklet; labels give the candidate, \pneo
$P(\mathrm{NEO})$, digest2 raw score, sky rate and band. The first panel
shows north and east.}
\label{fig:lsstcutouts}
\end{figure*}

The remaining, unlabeled tracklets also contain likely real moving objects with
no known orbit, including several that \pneo rates as likely NEOs
(Figure~\ref{fig:lsstcutouts}); a single night is not enough to establish
whether they are.

\begin{table}
\centering
\caption{Unassociated Rubin detections later matched to cataloged objects,
for arcs $\geq15$~min. Flagged counts are NEO / non-NEO / non-NEO with
$q\geq4.5$~AU.}
\label{tab:lsstunk}
\begin{tabular}{lcc}
\hline
 & 2+ detections & 3+ detections \\
\hline
Tracklets / NEOs & 3\,345 / 22 & 538 / 7 \\
\pneo All predicted & 22.8 & 4.8 \\
digest2 raw $\geq65$ & 20 / 29 / 17 & 7 / 2 / 2 \\
\pneo All $P\geq0.1$ & 21 / 14 / 0 & 7 / 1 / 0 \\
\pneo All $P\geq0.5$ & 16 / 2 / 0 & 4 / 0 / 0 \\
\hline
\end{tabular}
\end{table}

\subsection{Rubin Data Preview 2}
\label{sec:dp2}

Rubin Data Preview~2\footnote{\url{https://dp2.lsst.io}} \citep[hereafter
DP2;][]{RTN-115,dp2data} contains image and catalog products from the
processing of early
observations taken with the LSST Camera \citep[LSSTCam;][]{lsstcam} between
2025 April and 2026 January; it is the first release of LSSTCam data. Its
single-epoch images and catalogs cover
about 15\,000~deg$^2$, and its coadded images about 3\,000~deg$^2$. The
catalogs include the measurements made on difference images
(diaSources), the same kind of detections that the alert stream
distributes, with detections of known Solar System objects identified in
an SSSource table. The early release of 2026 July, which we use, contains
all catalogs but only coadded images. DP2 therefore lets us repeat the
alert-stream test on a larger, independently processed set of Rubin
detections.
\medskip

For this experiment, we use diaSources from the eight nights with data between 2025 July 3 and
11, processed as in Section~\ref{sec:lsstdata}. Known-object detections are
labeled with $q$ from the MPC orbit catalog, using the designations in
the SSSource table.

\paragraph{Known objects.}
The DP2 sample contains 285\,126 pure known-object tracklets, 393 of them
NEOs; \pneo predicts 399.3. For the 115\,906 tracklets with at least three
detections, it predicts 166.0 NEOs against 170 observed, and all 78 with
$P\geq0.9$ are NEOs. At its posting threshold of 65, digest2 selects 132 of
the 170 NEOs, together with 268 non-NEOs, 59 of them with $q\geq4.5$~AU; at
$P\geq0.07$, \pneo selects 132 NEOs and only 109 non-NEOs (about 60\% fewer), 2 of
them distant (Table~\ref{tab:dp2known}).

\begin{table}
\centering
\caption{Known objects in DP2 for pure tracklets with arcs
$\geq15$~min. ``Predicted'' is $\sum P$ for \pneo All. Flagged counts are
NEO / non-NEO / non-NEO with $q\geq4.5$~AU.}
\label{tab:dp2known}
\begin{tabular}{lcc}
\hline
 & 2+ detections & 3+ detections \\
\hline
Tracklets / NEOs & 285\,126 / 393 & 115\,906 / 170 \\
\pneo All predicted & 399.3 & 166.0 \\
digest2 raw $\geq65$ & 299 / 622 / 207 & 132 / 268 / 59 \\
\pneo All $P\geq P^{*}$$^a$ & 300 / 344 / 17 & 132 / 109 / 2 \\
\pneo All $P\geq0.1$ & 288 / 245 / 16 & 130 / 81 / 1 \\
\pneo All $P\geq0.5$ & 241 / 35 / 10 & 106 / 6 / 1 \\
\hline
\end{tabular}
\tablecomments{$^a$Largest threshold at which \pneo selects at least as many
NEOs as digest2 raw $\geq65$: $P^{*}=0.06$ (2+) and $0.07$ (3+).}
\end{table}

\begin{figure*}
\centering
\includegraphics[width=\textwidth]{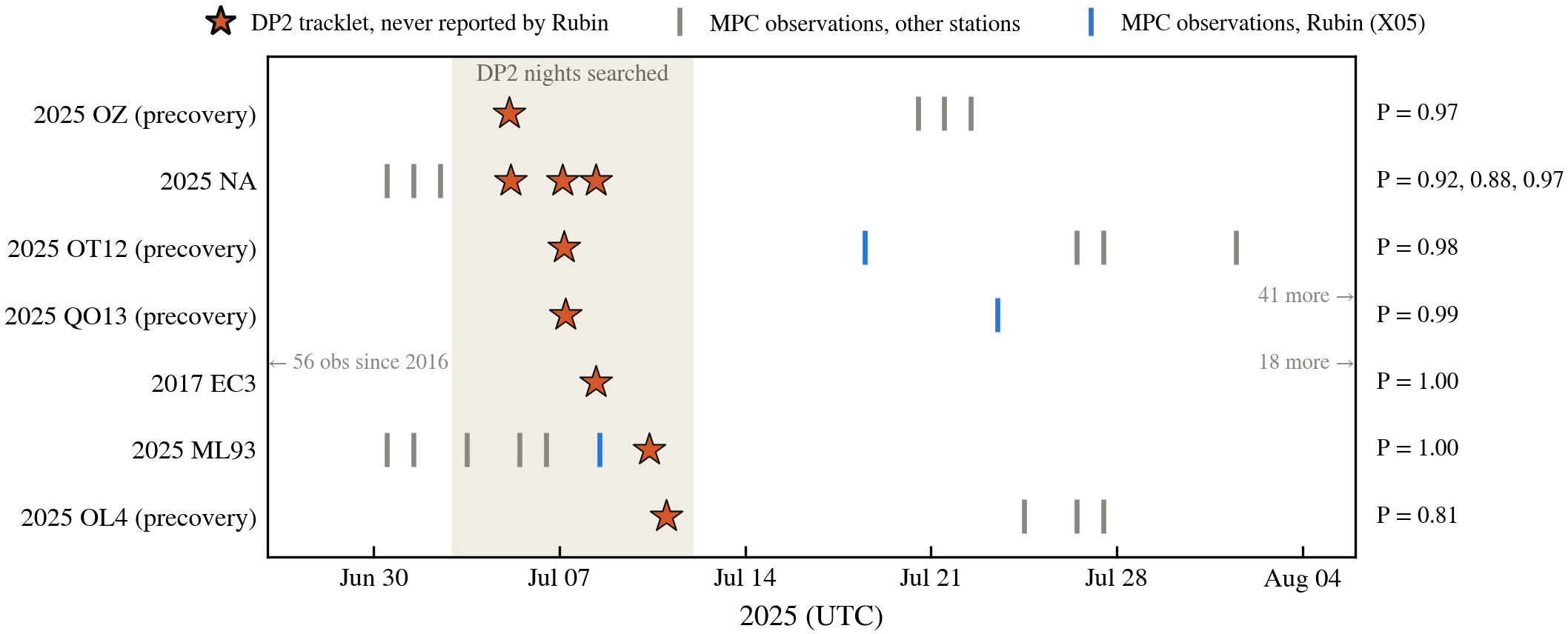}
\caption{Known NEOs among the DP2 candidates whose detections were not
reported to the MPC. Stars mark DP2 tracklets and give the \pneo probability;
ticks show MPC observations by night, with Rubin observations in blue. The
shaded region marks the DP2 nights searched.}
\label{fig:dp2precov}
\end{figure*}

\paragraph{Unreported NEOs.}
Among the unassociated tracklets with at least three detections, we look
for known objects that Rubin missed by propagating all 1.58 million MPC
asteroid orbits and 966 comet orbits with ASSIST \citep{holman2023assist} to their
detections. Sixteen are known objects that Rubin did not associate. Nine of
their tracklets belong to seven known NEOs, all assigned
$P(\mathrm{NEO})=0.81$--1.00 (Figure~\ref{fig:dp2precov}), and none of these
detections was submitted to the MPC. Four of the objects (2025~OZ,
2025~OT12, 2025~QO13 and 2025~OL4) were first reported to the MPC 11--16
days after the DP2 detections. A classifier applied to Rubin's unassociated
detections would therefore have flagged them as likely NEOs up to two weeks
earlier.

\subsection{Summary of the real-data tests}
\label{sec:rubinsummary}

The real-data tests confirm the simulation results. First, \pneo's
probabilities remain close to calibrated: the predicted number of NEOs
agrees with the observed number to within 2--4\% on the NEO Confirmation
Page, on Rubin detections of objects that Rubin failed to associate, and on
DP2. The one exception is the known objects in the alert stream, where
\pneo underpredicts by 15\%. The missing probability lies in slow-moving, distant
Amors that \pneo assigns low $P(\mathrm{NEO})$; a likely cause is that this
sample is selected by catalog membership, which enriches it in NEOs
relative to the population model (Section~\ref{sec:population}). Second, at the same NEO
recovery, \pneo selects far fewer non-NEOs than digest2: about 40\% fewer
on the NEO Confirmation Page, where digest2's input is already filtered,
and 50--75\% fewer on Rubin detections in both the alert stream and DP2, with almost none of the distant
objects that account for many of digest2's false alarms.

The remaining failures arise before classification. Pairs of unrelated
detections and repeated detections of static sources can form
plausible-looking tracklets that no classifier can reject from the tracklet
alone. Requiring at least three detections and removing repeat diaObjects
suppresses most of these cases.

\section{Discussion}
\label{sec:discussion}

\subsection{Why a Bayesian classifier}
\label{sec:bayes}

\pneo computes the posterior probability of NEO membership directly from
Bayes' theorem, with no tuned scores or separately chosen priors. Such a
posterior is provably optimal for classification: if the population model
is correct, no other method can miss fewer NEOs for the same amount of
telescope time spent on non-NEOs \citep{duda2001}. The guarantee also
degrades gracefully: the loss of performance is at most proportional to the
error in the computed probabilities \citep{devroye1996}, so a model that is
close to correct yields a classifier that is close to optimal. Our real-data
tests suggest that the present model is already close enough for this
purpose (Section~\ref{sec:rubinsummary}), and any improvement in the model
improves the classifier directly, without retuning
(Section~\ref{sec:limits}).

\subsection{Triage in the Rubin era}
\label{sec:lsstera}

Under the current posting criterion (digest2 score $\geq65$), more than 100
Rubin tracklets per night could qualify for the NEO Confirmation Page
during the first year of the survey, yet only 8\% of them would truly be NEOs
\citep{wagg2025}. \citet{wagg2025} propose reducing this load with
information beyond the tracklet; our approach instead improves the
classification score itself.

The gain is large. At the completeness of the NEOCP threshold, our
classifier selects about 96\% fewer non-NEOs than digest2 in simulation,
and 50--75\% fewer on Rubin detections, with almost none of the distant
objects that make up most of digest2's false positives
(Sections~\ref{sec:disagreement} and~\ref{sec:lsstunknown}). The smaller
gain on Rubin data is expected: its labeled samples consist of objects
already in the MPC catalog, which lacks most of the faint distant and
main-belt objects that dominate digest2's simulated false positives, and
real tracklets carry noise absent from the simulation. The same catalog
is why such contaminants are rare on today's NEO Confirmation Page. At
Rubin's depth, most of them will be undiscovered
(Section~\ref{sec:neocp}; \citealt{kurlander2025}), so the simulated gain
is the better guide to Rubin-era operations, within the uncertainty in
S3M's faint distant populations (Section~\ref{sec:limits}), and the gain
measured on Rubin data a conservative lower bound.

The gain matters most for tracklets with three or more detections, which
Rubin obtains where fields overlap. Most Rubin tracklets have only two
detections and must wait for multi-night linking, which can take up to
about two weeks \citep{wagg2025}. Tracklets with three or more detections
are rarely false pairings (12 of 208\,468 mixed tracklets in our alert
sample) and can be reported the same night. On these, both classifiers recover every known NEO, but \pneo's selection
is 86\% pure against 63\% for digest2, and its predicted NEO count agrees
with the observed one to within 10\% (Table~\ref{tab:lsstknown}), although
the sample contains only 12 NEOs.

Calibration also changes how follow-up can be planned. Summing
$P(\mathrm{NEO})$ over a night's candidates gives the expected number of
NEOs among them, so a threshold can be chosen by its expected yield and
contamination, and NEO follow-up can be weighed against other programs on
a common scale (Section~\ref{sec:comparison}), without an empirical mapping
from score to outcome for each survey. The strategies of \citet{wagg2025}
are complementary: whether Rubin is likely to recover a candidate on its
own, for example, is information beyond the tracklet that can be combined
with $P(\mathrm{NEO})$.

\subsection{Effective gain in follow-up capacity}
\label{sec:capacity}

Follow-up resources are limited, so what matters is the fraction of
follow-up targets that turn out to be NEOs. If half of the targets are NEOs
and a better classifier raises this to all of them, the same follow-up
effort recovers twice as many NEOs: the effective capacity of the follow-up
network doubles.

\pneo doubles this fraction in the Rubin-depth simulation, from 48\% to
96\% (Section~\ref{sec:comparison}), and raises it by a factor of 1.4--1.7
on Rubin detections (Sections~\ref{sec:lsstknown}--\ref{sec:dp2}). Even on
the NEO Confirmation Page, whose candidates have already been filtered by
digest2, it rises from 68\% to 78\%, a factor of 1.14
(Section~\ref{sec:neocp}).

The gain on Rubin data is likely a lower bound, because those samples are
drawn from cataloged objects (Section~\ref{sec:rubinsummary}). The
contaminants that \pneo removes are also mostly faint and distant, and
therefore expensive to observe, so the saving in telescope time may be
larger than these ratios suggest. We therefore expect \pneo to potentially
double the effective capacity of the follow-up network in the Rubin era.
The time saved could go to fainter or more numerous candidates, or to
higher completeness.

\subsection{What a single tracklet cannot tell}
\label{sec:lowp}

Some NEOs cannot be distinguished from main-belt asteroids in a single
night. For these, \pneo assigns low or intermediate probabilities rather
than forcing a distinction the data do not support, so its performance is
ultimately limited by the size of the undiscovered main-belt population.
This limit will ease as the survey proceeds: within its first two to three
years, LSST will have discovered most of the main-belt asteroids within its
reach \citep{wagg2025}, which can then be identified as known objects before
classification.
These low probabilities are themselves calibrated: in the self-consistency
samples, 540 NEOs receive $P(\mathrm{NEO})<0.1$, against 535.7 expected.
They move slowly (median $0.41^\circ$\,day$^{-1}$, against
$1.5^\circ$\,day$^{-1}$ for all NEOs), are observed at larger distances
(median 1.27 against 1.04~AU), and lie preferentially near the NEO boundary:
one quarter of NEOs with $1.2\leq q<1.3$~AU receive $P<0.1$, while no NEO
moving faster than $1^\circ$\,day$^{-1}$ does. In the Rubin alerts, the
low-probability NEOs are likewise mostly distant Amors
(Section~\ref{sec:lsstknown}).

\subsection{Limitations and Future Work}
\label{sec:limits}

The population model is the main remaining uncertainty, but its effect is
narrower than its known errors might suggest. \pneo remains close to
calibrated on real data despite S3M's overprediction of some bright
populations (Section~\ref{sec:popcaveats}), its coarse $(q,e,i,H)$
histogram with uniform angular elements, and its support only to $H=30$.
In hindsight, this is not surprising: most tracklets are unambiguous, so their probabilities
are near 0 or 1 under any reasonable model, and $P(\mathrm{NEO})$ depends
only on the ratio of the two class densities where a tracklet's solutions
fall; for ranking and triage, a rough model is therefore sufficient.

The
model matters for tracklets with intermediate probabilities, such as
Mars-crossers and slow distant Amors near the NEO boundary, where an error
in the local ratio of NEOs to main-belt asteroids changes the odds
directly, and it sets the size of the simulated gain over digest2, which
depends on S3M's poorly constrained faint distant populations. Our
real-data tests, which compare total counts, constrain these effects only
weakly. Because the population enters the classifier only through its
density (Section~\ref{sec:implementation}), a newer model, such as
NEOMOD3 \citep{nesvorny2024} for the NEOs or a future one inferred directly from Rubin discoveries, can
be substituted without changing the inference framework; testing how
the probabilities respond to such substitutions is a natural next step. As
Rubin catalogs more of the main belt, updating the Unk population lets the
classifier follow the shrinking discovery background without retuning.

Two further limitations concern the input tracklet rather than the
classifier. The first is measurement error: \pneo treats the attributable
as exact, which is adequate for arcs of 15~minutes or longer
(Section~\ref{sec:robustness}), but 30\% of same-night arcs in Rubin's
baseline cadence are shorter. For these, the astrometric uncertainty should
be marginalized explicitly (Section~\ref{sec:noise}), which the software
already supports. The second is gross errors (mismatches) in constructing the tracklet. These must be removed upstream, for example by
requiring three or more detections and removing repeat diaObjects
(Section~\ref{sec:lsstcut}).

Finally, the present model contains only NEO and non-NEO classes, excludes
comets and interstellar objects, and treats each night independently.
Extending the framework to additional Solar System populations and to
multi-night observations are natural next steps.

\section{Summary}
\label{sec:summary}

We have presented \pneo, a Bayesian classifier for single-night tracklets.
It computes the posterior probability that a tracklet belongs to an NEO
directly from Bayes' theorem, by integrating an absolute population model
over all orbits consistent with the tracklet, with no tuned scores or
separately chosen priors. Our main results are:

\begin{enumerate}
    \item On identical simulated inputs to depths expected of steady-state Rubin/LSST, at the completeness of the current
    NEOCP posting threshold, \pneo selects 26 times fewer false positives (non-NEOs) than
    digest2, doubling the purity of the selection (96\% against 48\%).
    Three-quarters of digest2's false positives are distant objects
    ($q\geq4.5$~AU), whose slow tracklets also admit nearby, NEO-like
    solutions; our posterior correctly gives the distant solutions far more
    weight, and \pneo rejects them.

    \item On real data, \pneo's predicted NEO counts agree with those observed
    to within 2--4\% on the NEO Confirmation Page, on Rubin detections that
    Rubin failed to associate, and on Rubin Data Preview~2. At the same completeness, \pneo selects
    1.6 times fewer non-NEOs than digest2 on the NEO Confirmation Page, whose
    candidates have already been filtered by digest2, and 2--4 times fewer on
    Rubin detections in both the alert stream and Data Preview~2, with almost no distant contaminants.
    In Data Preview~2 it assigns $P(\mathrm{NEO})\geq0.81$ to seven known NEOs
    that Rubin detected but did not report, four of them 11--16 days before
    they were first reported to the MPC.

    \item \pneo is calibrated on its own population model: on 800\,000
    simulated tracklets, the predicted number of NEOs agrees with the
    observed number to 0.1\%, with an odds scale of $1.01$ $[0.94,1.08]$.

    \item Because \pneo computes a true Bayesian posterior, its
    classifications are optimal for the adopted population model, and remain
    close to optimal for any model that is close to correct
    (Section~\ref{sec:bayes}).

    \item The probabilities are robust to Rubin-like astrometric errors for
    arcs of at least 15~minutes, which covers about 70\% of Rubin tracklets.

    \item The implementation scores one tracklet in about 12~ms on a single
    core, or about $1.5\times10^{7}$ tracklets per hour on one server, allowing
    for application at Rubin scale.

\end{enumerate}

For follow-up in the era of large surveys, the main advantages are a much
cleaner candidate list, particularly for the tracklets with three or more
detections that can be reported the same night, and a priority with a
direct probabilistic interpretation: follow-up programs can choose targets
by expected NEO yield and weigh them against other observations on a common
scale, rather than relying on a survey-dependent ranking score.
In the Rubin era, the cleaner list could potentially double the
effective capacity of the follow-up network (Section~\ref{sec:capacity}). A classifier
of this kind transforms NEO follow-up in the era of deep surveys: it turns
an overwhelming stream of candidates into a short list of probable NEOs,
and with it makes effective follow-up possible at the depth of Rubin.

\begin{acknowledgments}
We thank the DiRAC solar system group for many helpful conversations. D.S. thanks Lara Heber for motivating discussions for this project.

D.S. and M.J. acknowledge support from the DiRAC Institute in the Department of Astronomy at the University of Washington. The DiRAC Institute is supported through generous gifts from the Charles and Lisa Simonyi Fund for Arts and Sciences, Janet and Lloyd Frink, and the Washington Research Foundation.

M.J. acknowledges support from NASA grant 80NSSC24K0852, awarded under the Yearly Opportunities for Research in Planetary Defense (YORPD) program.

This research has made use of data and services provided by the International Astronomical Union’s Minor Planet Center, and of the JPL Small-Body Database. We used alert data from the Fink and ALeRCE brokers, and data products from NSF–DOE Vera C. Rubin Observatory.
This publication is based in part on proprietary Rubin Observatory Legacy Survey of Space and Time (LSST) data, and was prepared in accordance with the Rubin Observatory data rights and access policies. All authors of this publication meet the requirements for co-authorship of proprietary LSST data. This research uses services or data provided by the Rubin Science Platform at NSF-DOE Vera C. Rubin Observatory, which is jointly funded by the U.S. National Science Foundation and the U.S. Department of Energy, Office of Science.

The authors used AI coding assistants
(Claude, Anthropic) to help develop software and analysis scripts and to write
and edit the manuscript. The authors reviewed, tested, and take full responsibility for
all code, results, and text.

\end{acknowledgments}

\begin{contribution}

D.S.\ carried out the validation and analyses, extending the initial
comparison with digest2 and conducting the NEO Confirmation Page, Rubin
alert-stream, and Data Preview~2 tests. D.S.\ developed the software for
these tests and led the writing of the manuscript. M.J.\ conceived the project and the classifier algorithm, led the
implementation, and advised D.S.
\v{Z}.I.\ advised D.S.\ and verified the analysis.

\end{contribution}

\facilities{Vera C. Rubin Observatory, Minor Planet Center, JPL Small-Body Database Query}

\software{\pneo, digest2 \citep{keys2019,veres2023}, HelioLinC \citep{heinze2022AAS},
ASSIST \citep{holman2023assist}, REBOUND, Sorcha \citep{merritt2025},
rubin\_sim, Astropy \citep{astropy2022}, NumPy, SciPy, Matplotlib,
jplephem, PyO3, Rayon}

\appendix

\section{The change-of-measure Jacobian}
\label{app:jacobian}

\paragraph{Observable block.}
Order the inputs as $(\alpha,\delta,\rho\,|\,\dot\alpha,\dot\delta,\dot\rho)$
and the outputs as $(\mathbf{r}\,|\,\mathbf{v})$; reordering columns changes
only the sign of the determinant. By Equation~(\ref{eq:r}), $\mathbf{r}$ does
not depend on the three rates, so the matrix is block triangular,
\begin{equation}
\frac{\partial(\mathbf{r},\mathbf{v})}
     {\partial(\alpha,\delta,\rho,\dot\alpha,\dot\delta,\dot\rho)}
=\begin{pmatrix} P & 0\\ S & P\end{pmatrix},
\quad
P=\left[\rho\hat{\mathbf{u}}_\alpha\;\;\rho\hat{\mathbf{u}}_\delta\;\;
\hat{\mathbf{u}}\right],
\end{equation}
and its determinant is $(\det P)^2$; the lower-left block $S$, which holds the
second derivatives of $\hat{\mathbf{u}}$ and all dependence on the rates,
does not enter. The vectors $\hat{\mathbf{u}}_\alpha$,
$\hat{\mathbf{u}}_\delta$ and $\hat{\mathbf{u}}$ are mutually orthogonal with
lengths $\cos\delta$, 1 and 1, so $|\det P|=\rho^2\cos\delta$ and
\begin{equation}
\left|\det\frac{\partial(\mathbf{r},\mathbf{v})}
{\partial(A,\rho,\dot\rho)}\right| = \rho^4\cos^2\delta.
\label{eq:obsblock}
\end{equation}
Two powers of $\rho$ come from converting angular position into physical
position and two from converting angular rates into transverse velocity;
$\cos^2\delta$ reflects the convergence of lines of constant right ascension
toward the poles. The equatorial-to-ecliptic rotation between the two blocks
is orthogonal and does not change the determinant.

\paragraph{Element block.}
The Delaunay variables $(L,G,H_{\rm D},l,g,h)$, with $L=\sqrt{\mu a}$,
$G=L\sqrt{1-e^2}$, $H_{\rm D}=G\cos i$ and $(l,g,h)=(M,\omega,\Omega)$, are
canonical, so the map from $(\mathbf{r},\mathbf{v})$ to them preserves
phase-space volume and has unit determinant. The angles map onto
$(M,\omega,\Omega)$ unchanged, and $(L,G,H_{\rm D})$ depend on $(a,e,i)$
triangularly, with diagonal terms $\partial L/\partial a=\tfrac12\sqrt{\mu/a}$,
$\partial G/\partial e=-Le/\sqrt{1-e^2}$ and
$\partial H_{\rm D}/\partial i=-G\sin i$. Their product gives
$|\det\partial(L,G,H_{\rm D})/\partial(a,e,i)|=\mu^{3/2}\sqrt{a}\,e\sin i/2$,
and hence
\begin{equation}
\left|\det\frac{\partial E}{\partial(A,\rho,\dot\rho)}\right|
= \rho^4\cos^2\delta\,\frac{2}{\mu^{3/2}\sqrt{a}\,e\sin i}
\label{eq:delaunay}
\end{equation}
for $E=(a,e,i,\Omega,\omega,M)$, exactly. Replacing $a$ by $q=a(1-e)$ at fixed
$e$ multiplies volumes by $\partial q/\partial a=1-e$, which gives
Equation~(\ref{eq:jacqei}).

\paragraph{Verification.}
The implementation evaluates Equation~(\ref{eq:jacqei}) in closed form. An
independent reference implementation instead builds the full $6\times6$
Jacobian from the analytic partial derivatives of each step and takes its
determinant numerically; the two agree to a maximum relative difference of
$1.2\times10^{-11}$ (Section~\ref{sec:implementation}).

\paragraph{The $1/(e\sin i)$ factor.}
For a circular orbit the perihelion direction, and so $\omega$, is
undefined, and for a zero-inclination orbit so is $\Omega$, just as longitude
is undefined at a pole; a small cell of states is then spread over a wide
range of elements. The singularity is integrable, so $\Lambda_C$ stays
finite. A smooth population vanishes as $e\to0$ and $i\to0$ and cancels it,
but the binned population model density is constant across its lowest bins
($e<0.1$, $i<2^\circ$), which affects the quadrature
(Appendix~\ref{app:quadrature}).

\paragraph{Relation to previous work.}
\citet[their \S2.3]{peiris2003} give the element-space identity by the same
Delaunay argument \citep[see also][]{tremaine2020}, and \citet{masuda2026}
derive the same closed form to transform Bayesian priors between Cartesian and
orbital coordinates for exoplanets. We have not found the composition with
$\rho^4\cos^2\delta$ used before for short-arc population classification.
\citet[][\S3.2]{farnocchia2015}, mapping an S3M population onto the
$(\rho,\dot\rho)$ plane for Scout, call the transformation ``quite
complicated'' and use the position factor $\rho^2$ alone, and
\citet{hussein2014} judge the equivalent transformation ``impossible to
achieve in closed-form'' and evaluate it by Monte Carlo.

\section{Admissible region and quadrature}
\label{app:quadrature}

\paragraph{The $\dot\rho$ interval.}
Write Equation~(\ref{eq:v}) as $\mathbf{v}=\mathbf{w}+\dot\rho\,\hat{\mathbf{u}}$,
where $\mathbf{w}=\dot{\mathbf{R}}+\rho(\dot\alpha\,\hat{\mathbf{u}}_\alpha+
\dot\delta\,\hat{\mathbf{u}}_\delta)$ is fixed once $\rho$ is chosen. The
bound-orbit condition is then quadratic in $\dot\rho$ and holds for
\begin{equation}
\bigl|\dot\rho+\hat{\mathbf{u}}\cdot\mathbf{w}\bigr|
<\Bigl(\frac{2\mu}{r}-|\mathbf{w}_\perp|^2\Bigr)^{1/2},
\label{eq:rhodot}
\end{equation}
where $\mathbf{w}_\perp$ is the part of $\mathbf{w}$ perpendicular to the
line of sight. The interval is empty when the right-hand side is imaginary.
Since $|\mathbf{w}_\perp|$ grows with $\rho$, this also limits how far away a
fast-moving tracklet can be. The $\rho$ cell edges lie at the geometric
midpoints between neighboring grid points, and columns with an empty interval are
dropped.

\paragraph{Why the grid is finer in $\rho$.}
The binned density jumps at every bin edge. Edges in $q$, $e$ and $i$ cross
the admissible region in both directions, but because $H$ depends on $\rho$
and not on $\dot\rho$, every $H$-bin edge is a line of constant $\rho$ that
cuts each cell of its column at the same place. Each grid point stands for its
whole cell, so where an edge cuts a cell the grid point's value is wrong for part of
it. For an edge crossing the grid at a slant these errors change sign from
cell to cell and partly cancel, and refining either axis reduces them. For an
$H$ edge they are identical in every row and add coherently: refining
$\dot\rho$ leaves them unchanged, and only narrower columns reduce them. At
equal cost, a $(n_\rho,n_{\dot\rho})=(800,100)$ grid has 2.3 times smaller
error than $(400,200)$, while $(200,400)$ is no better than $(200,100)$. At
the default grid this staircase is no longer the dominant error. The
$1/(e\sin i)$ factor (Appendix~\ref{app:jacobian}) adds narrow, integrable
spikes along the curves $e=0$ and $i=0$, across which grid quadrature
converges slowly.

\paragraph{Accuracy.}
The reference grid of $9600\times400$ changes by $5\times10^{-4}$ (90th
percentile) when refined by a further factor of two in each direction. For
small relative errors $\delta_C$ in the $\Lambda_C$, the posterior changes by
$\Delta P\approx P(1-P)(\delta_{\rm NEO}-\delta_{\rm other})$, so tracklets
with $P$ near 0 or 1 show almost no error however poorly their $\Lambda_C$ are
resolved. On a realistic mixture, dominated by $P\approx0$, the
unconditional 90th-percentile error is $\sim5\times10^{-5}$ whatever the
quadrature does, which is why we quote only $0.05<P<0.95$. For those 497
tracklets the 90th percentile is $8\times10^{-3}$. The largest single error is
0.23; a maximum over a few hundred tracklets is set by the one worst-resolved
case, so we do not use it as a summary.

\section{The $H$ extension and the triage policy}
\label{app:truncation}

\paragraph{Where the $H$ boundary binds.}
An object at heliocentric distance $r$ is brightest at opposition, at a
distance of about $r-1$~AU from the observer and zero phase angle, so a
survey reaching $V_{\lim}$ detects objects up to
\begin{equation}
H_{\max}(r)\approx V_{\lim}-5\log_{10}\bigl[r(r-1)\bigr].
\label{eq:hmax}
\end{equation}
For $V_{\lim}=24$:
\begin{center}
\begin{tabular}{lcccccc}
\hline
$r$ (AU) & 1.3 & 1.45 & 1.55 & 1.67 & 1.8 & 2.0 \\
$H_{\max}$ & 26.0 & 24.9 & 24.3 & 23.8 & 23.2 & 22.5 \\
\hline
\end{tabular}
\end{center}
Main-belt orbits ($r\gtrsim1.66$~AU) cannot reach $H=25.5$; only NEOs and
Mars-crossers observed near a perihelion just outside $q=1.3$~AU can. With the
extension, the $H=30$ boundary can bind at $V\le24$ only for hypotheses with
$\rho r\lesssim0.06$~AU$^2$, i.e.\ within about 0.06~AU of the observer.

\paragraph{Why $H=30$.}
At 30, four of the 11 NEOCP boundary tracklets keep 7--10\% of their
admissible measure above the boundary, which leaves their $P$ unchanged
because $\Lambda_{\rm other}=0$ for all four. Extending to 34 would cover the
faintest synthetic tracklets of Section~\ref{sec:validation} (maximum implied
$H=33.8$), but would place 76\% of the NEO population in the synthetic bin.

\paragraph{The triage policy.}
The policy fires only when the admissible region is non-empty, both
$\Lambda_C$ are exactly zero, and at least 95\% of the quadrature measure of
grid points with finite implied $H$ lies at or above the support boundary. It then
assigns $P(\mathrm{NEO})=1$, leaves $\ell$ undefined, and flags the result.
The reasoning is physical as well as empirical: a tracklet too faint for the
model at nearly every admissible distance can only be a nearby, small object,
and all 11 such NEOCP tracklets were NEOs.
The 95\% threshold lies in a measured gap: the 11 boundary tracklets have
fractions of at least 0.986, and every constructed counter-example, such as a
bright, slow tracklet whose admissible region lies in empty cells of the
model, has a fraction of exactly zero. Its evidence is thin: 11 objects,
all from the sample that prompted it.

\section{Rubin alert-stream processing}
\label{app:rubindata}

\paragraph{Data.}
We downloaded alerts through the Fink broker's Data Transfer service
\citep{moller2021} for 2026 July 5--13 and October 29--30. A June 30 night
included in the transfer is excluded. The unfiltered transfer contains
3\,789\,349 alerts: 291\,305 associated by Rubin with known Solar System
objects, 2\,141\,791 on newly created diaObjects, and 1\,356\,253 on
previously existing diaObjects (Rubin's records of individual sources seen
in difference images; \citealt{LSE-163}).

\paragraph{Known-object detections.}
The associated alerts cover 113\,687 objects. After removing the June night
and 17 detections with non-positive flux, 288\,728 detections remain.
Magnitudes are computed from the PSF flux as
$m=31.4-2.5\log_{10}(F/\mathrm{nJy})$ in the Rubin band and converted to
Johnson $V$ using the digest2 color table.

\paragraph{Unassociated detections.}
For detections without a known-object association, we require positive flux,
no dipole, glint, or streak flag, and real/bogus reliability $\geq0.9$.
These cuts leave 291\,177 detections: 190\,192 on new diaObjects and
100\,985 on repeat diaObjects. The reliability threshold retains 95.4\% of
the known-object detections.

Repeat diaObjects are often variable stars, nuclear transients, or persistent
artifacts. We therefore remove detections whose diaObject has more than two
detections ($n_{\rm DiaSources}>2$), leaving 217\,663 detections.

\paragraph{Tracklets.}
We link detections blindly into single-night tracklets with HelioLinC's
\texttt{make\_tracklets} \citep{heinze2022AAS}. Two detections on
overlapping exposures (assumed image radius $2^\circ$) are paired if they
are at most 1.5~h apart and imply a sky motion of at most
$5^\circ\,\mathrm{day}^{-1}$; pairs are extended into longer tracklets when
all detections lie within $0.5''$ of a great-circle fit. Increasing the speed limit to
$10^\circ\,\mathrm{day}^{-1}$ adds no correctly linked tracklets in the known
sample and only increases links between unrelated objects. Object identities,
where available, are attached only after linking.

Among known objects, 78\,657 of 79\,048 object-nights with at least two
detections (99.5\%) are recovered as pure tracklets. For NEOs the recovery is
111 of 112; the missed object has only two detections separated by 2.3~h. We
retain arcs of at least 15~minutes, the shortest duration for which the classifier
remains calibrated in Section~\ref{sec:robustness}.

Because \texttt{make\_tracklets} has no minimum speed, repeated detections of
a stationary source can link to one another. We remove tracklets that move less than $0.5''$ in total. These have median
motion $0.07''$, compared with $0.65''$ for cataloged asteroids whose
detections Rubin placed on a single diaObject. A more aggressive $2''$ cut
would also remove 35 distant asteroids ($q\geq4.5$~AU), so we do not use it.

\paragraph{Labels.}
Known-object tracklets are labeled from their MPC designation and the MPC
orbit catalog \citep[MPCORB, version of 2026 November 2;][]{mpcorb}, with NEOs defined by $q<1.3$~AU. All
but one of the 113\,687 Rubin-associated objects match the catalog; 213 are
NEOs. We call a tracklet \emph{pure} when all detections belong to the same
object and \emph{mixed} otherwise.

For unassociated tracklets, we propagate all 1\,578\,692 MPC orbits from the
2026 November 2 catalog to the time of each detection, including light time
and Rubin's topocentric position. On known detections, the propagated
positions agree with the observations to 0.2--$1.3''$. We assign a label only
when every detection in a tracklet lies within $10''$ of the same object,
which suppresses chance single-detection matches. Rubin observations from
these nights had not yet reached the MPC when this matching was performed and
therefore cannot be used as labels.

\paragraph{Scoring.}
Each tracklet is written to ADES format using observatory code X05, with
times converted from TAI to UTC. We score every tracklet with \pneo using both
the All and Unk priors and with digest2 2.8.2 using its raw and NoID scores.
Every tracklet receives all four scores.

\bibliography{bibliography}{}
\bibliographystyle{aasjournalv7}

\end{document}